\documentclass{elsart}
\def\Journal#1#2#3#4{{#1} {\bf #2} (#4) #3}

\def\NIMA{{\em Nucl. Instrum. Methods} A}

\def\CRRC2{$CRRC^2$}
\newcommand{\promille}{%
  \relax\ifmmode\promillezeichen
        \else\leavevmode\(\mathsurround=0pt\promillezeichen\)\fi}
\newcommand{\promillezeichen}{%
  \kern-.05em%
  \raise.5ex\hbox{\the\scriptfont0 0}%
  \kern-.15em/\kern-.15em%
  \lower.25ex\hbox{\the\scriptfont0 00}}
\usepackage{epsfig}

\usepackage{amssymb}

\newcommand{\LAr}          {liquid argon}
\newcommand{\Ical}         {I^\mathrm{cal}}
\newcommand{\Iion}         {I^\mathrm{ion}}
\newcommand{\Td}           {T_\mathrm{D}}
\newcommand{\Tc}           {\tau_\mathrm{c}}
\newcommand{\gcal}         {g^\mathrm{cal}}
\newcommand{\gion}         {g^\mathrm{ion}}
\newcommand{\Exp}[1]       {\mathrm{e}^{#1}}

\newcommand{\invLapl}[1]   {{\mathcal L}^{-1}\left[{#1}\right]}

\begin{document}

\begin{frontmatter}



\title{Response Uniformity of the ATLAS Liquid Argon Electromagnetic Calorimeter}


\author[Annecy]{M.~Aharrouche\thanksref{ElKacimi}},
\author[Annecy]{J.~Colas},
\author[Annecy]{L.~Di Ciaccio},
\author[Annecy]{M.~El~Kacimi\thanksref{ElKacimi}},
\author[Annecy]{O.~Gaumer\thanksref{Gaumer}},
\author[Annecy]{M.~Gouan\`ere},
\author[Annecy]{D.~Goujdami\thanksref{ElKacimi}},
\author[Annecy]{R.~Lafaye},
\author[Annecy]{S.~Laplace},
\author[Annecy]{C.~Le Maner\thanksref{Maner}},
% ...

\address[Annecy]{Laboratoire de Physique de Particules (LAPP), 
Universit\'e de Savoie, CNRS/IN2P3, Annecy-le-Vieux~Cedex, France.} 
% ...

\thanks[Gaumer]{Now at university of Geneva, switzerland.}
\thanks[Prieur]{Now at Rutherford Appleton Laboratory (RAL), Chilton, Didcot, OX11 0QX, United Kingdom.}
\thanks[Maner]{Now at university of Toronto, Ontario, Canada.}
\thanks[ElKacimi]{Visitor from LPHEA, FSSM-Marrakech (Morroco).}
\thanks[Deceased]{Deceased.}
\thanks[Toth]{Also at KFKI, Budapest, Hungary.
  Supported by the MAE, France, the
  HNCfTD (Contract F15-00) and the Hungarian OTKA (Contract T037350).}
\thanks[cauthor]{Corresponding author, E-mail: kado@lal.in2p3.fr}
\thanks[Sacco]{Now at Queen Mary, University of London.} 
\thanks[Poggioli]{Now at LAL, Univ Paris-Sud, IN2P3/CNRS, Orsay, France.}
\thanks[Benchouk]{Now at Facult\'e de Physique, Universit\'e des Sciences et de la technologie Houari Boumedi\`ene, BP 32 El-Alia 16111 Bab Ezzouar, Alger, Alg\'erie.} 
\thanks[Unal]{Now at CERN, Geneva, Switzerland.} 


\begin{abstract}

The construction of the ATLAS electromagnetic liquid argon calorimeter
modules is completed and all the modules are assembled and inserted in
the cryostats. During the production period four barrel and three
endcap modules were exposed to test beams in order to assess
their performance, ascertain the production quality and
reproducibility, and to scrutinize the complete energy
reconstruction chain from the readout and calibration electronics to
the signal and energy reconstruction. It was also possible to
check the full Monte Carlo simulation of the calorimeter. The analysis
of the uniformity, resolution and extraction of constant term is
presented. Typical non-uniformities of 5~\promille\ and typical
global constant terms of 6~\promille\ are measured for the barrel and
end-cap modules. 
\end{abstract}

\end{frontmatter}



\section*{INTRODUCTION} 
\label{Sec:Introduction}

The large hadron collider (LHC) will collide 7 TeV proton beams with
luminosities ranging from 10$^{33}$ to 10$^{34}$cm$^{-2}$s$^{-1}$. The
very high energy and luminosity foreseen will allow to investigate the
TeV scale in search for new phenomena beyond the Standard
Model. Reaching such performance is an outstanding challenge for both
the collider and the detectors.

The electromagnetic calorimeter of the ATLAS detector, one of the two
multi purpose experiments at LHC, is a lead and liquid argon sampling
calorimeter with accordion shaped absorbers. In its dynamic range
covering the few GeV up to the few TeV domain an excellent measurement
of the energy of electrons and photons is required in order to resolve
potential new particle resonances, and to measure precisely particle
masses and couplings.

One of the salient benchmark processes that has guided the design of
the electromagnetic ATLAS calorimeter is the Higgs boson production
with subsequent decay into a pair of photons. This event topology will
be observable only if the Higgs boson mass is smaller than
$\sim$150~GeV/c$^2$. In this channel the capabilities of the
calorimeter in terms of photon pointing resolution and
$\gamma$/$\pi^0$ discrimination are of chief concern. An excellent and
uniform measurement of the photon energy is essential. Another process
involving the Higgs boson where it decays to a pair of Z bosons and
subsequently into four electrons also requires a uniform measurement
of the electron energy over a large dynamic range. Finally, among the
processes which will require the most accurate knowledge of the
electron energy is the W mass measurement whose goal
precision is $\sim$0.2\promille. For all these processes the
constant term $b$ of the three main resolution elements:
$$\frac{\sigma_E}{E} = \frac{a}{\sqrt{E}} \oplus b \oplus
\frac{c}{E}$$ (where $c$ and $a$ are the noise and stochastic terms
respectively) plays an important role. It arises from various sources
encompassing the calibration and readout electronics, the amount of
material before and in the calorimeter, the energy reconstruction
scheme and its stability in time. The $c$ term comprises the
electronic noise and the pile-up.
 
Many years of R\&D work~\cite{RD3_1,RD3_2,RD3_3,RD3_4,RD3_5} have led
to the ATLAS calorimeter design whose first performance assessments on
pre-production modules were reported
in~\cite{NIMBARREL0,NIMENDCAP0}. Since then, modifications were made
in order to improve the production, the quality control and the
performance of the calorimeter modules. During the three years of
module construction, four barrel and three endcap modules have been
exposed to electron beams in the North Area at CERN's SPS. The primary
aim was to assess the quality of the production by measuring the
response uniformity over the complete acceptance. However, these
measurements have led to numerous further improvements of the calibration, signal reconstruction and the simulation of the calorimeter. 

This paper reports on measurements of the uniformity and a study of
the different contributions to the constant term of the electron
energy resolution for barrel and endcap modules exposed to high
energy electron beams. The actions taken to optimize the electron
energy resolution and in particular the uniformity of the response are
also described. Finally a review of all sources of
non-uniformities is presented.

The paper is organized as follows. In Sec.~\ref{Sec:Calomod} the main
features of the calorimeter modules are described and the differences
with respect to the pre-production modules are briefly presented. A
description of the readout and calibration electronics is then given
in Sec.~\ref{Sec:SignalRec}. The signal reconstruction, including
cross talk issues is also presented. The beam test setups are
described in section~\ref{Sec:TBsetup}. Finally the barrel and endcap
analysis and results are presented in Sec.~\ref{Sec:BarrelUnif} and
Sec.~\ref{Sec:Endcap}.

\newcommand{\degr}{\mbox{$^\circ$}}

\section{CALORIMETER MODULES DESCRIPTION} 
\label{Sec:Calomod}

The ATLAS electromagnetic calorimeter is a lead and liquid argon
sampling calorimeter using an accordion geometry for gaps and
absorbers. It is composed of a cylindrical barrel centered on the beam
and two endcaps. The usual slightly modified polar coordinate
system is used, where the $z$-axis coincides with the beam axis,
$\phi$ is the azimuthal angle and the polar angle $\theta$ is replaced
by the pseudo rapidity $\eta=-ln (tan\theta/2)$. A detailed
description of the detector can be found
in~\cite{TDR-LARG,Barrel-construction,Electrodes,PSpaper}. The main
characteristics are, however, described hereafter.

\subsection{\bf Barrel Modules Description}
\label{BarrelMod}
The ATLAS barrel calorimeter is divided in two half barrels covering
the positive and the negative pseudo rapidities, from $|\eta| = 0 $ to
$|\eta| = 1.475$. The inner and outer diameters are about 2.8 m and 4
m. Each half barrel consists of sixteen 3.2 meter long modules made of
64 accordion shaped absorbers interleaved with readout
electrodes. Modules are made of two types of electrodes denoted A and
B covering the pseudo rapidity ranges $[0,0.8]$ and $[0.8,1.475]$
separated by 2.5 mm at the transition point.  The construction and
testing of these modules is reported in~\cite{Barrel-construction}. In
each of these regions the lead thickness is different in order to
maintain an approximately constant stochastic term of the energy
resolution (1.53 mm for $|\eta| <0.8$ and 1.13 mm for
$|\eta|>0.8$). The drift gap on each side of the electrode is 2.1 mm
which corresponds to a total drift time of about 450 ns at 2~kV.  The
electrodes are segmented in three compartments in depth (front,
middle, back). The first section with narrow strips along $\eta$, ends
after 4 radiations lengths ($X_0$), while the second stops after 22
$X_0$. The total thickness of the module varies as a function of
$\eta$ between 22 and 33 $X_0$.  Summing boards, plugged on the
electrodes, perform the signal summation in $\phi$~: 16 adjacent
electrodes are summed for a strip cell and 4 electrodes for middle and
back cells. The granularity of the different compartments are shown in
Table~\ref{tab:granularity} for $|\eta|<1.475$. A presampler is placed
in front of the electromagnetic calorimeter inside the cryostat. Each
unit provides coverage in $|\eta|$ from 0 to 1.52 and covers a region
of $\Delta\phi=0.2$.  The signal is sampled in a thin active layer of
11~mm of liquid argon with a readout cell granularity of $\Delta\eta
\times \Delta\phi = \sim0.025 \times 0.1$~\cite{PSpaper}. In total,
there are 3424 readout cells per module. The high voltage is supplied
independently on each side of the electrode by sectors of $\Delta \eta
\times \Delta \phi$=0.2$ \times$ 0.2 corresponding to 32 electrodes.

 
Four barrel modules (two of each half barrel) where exposed in the H8
beam test line, namely M13, M10, P15 and P13. They are now part of the
final ATLAS calorimeter. However, because of the incorrect mounting of
the support of a cerenkov counter, non homogeneous material was added
in the beamline while the module M13 was being tested. As a
consequence its performance could not be thoroughly
measured. Therefore, only results for M10, P13 and P15 are presented.

\subsection{\bf Endcap Modules Description}

The ATLAS electromagnetic endcap calorimeter (EMEC) covers the
rapidity region from 1.375 up to 3.2 and consists of 2 wheels, one on
each side of the electromagnetic barrel.  Each wheel, divided into
eight wedge-shaped modules, is 63 cm thick with internal and external
radii of about 30~cm and 2~m.  For technical reasons, it is divided in
one outer ($1.375<|\eta|<2.5$) and one inner ($2.5<|\eta|<3.2$) wheel,
both projective in $\eta$.  The crack between the wheels is about 3~mm
wide and is mainly filled with low density material.  In $\phi$, one
outer (inner) wheel module is made of 96 (32)~accordion shaped
absorbers interleaved with readout electrodes, covering a range of
$\Delta\phi=2\pi/8$.  Accommodating the accordion shape in the endcap
region induces a quite complicated geometry.  Most of the design
parameters vary along the radius direction (corresponding to $\eta$)~:
liquid argon gap, sampling fraction, accordion wave amplitudes or
folding angles.  A continuously varying high voltage setting along
$\eta$,
would partially compensate for this and imply an almost
$\eta$-independent detector response.  However,
the high voltage is set by steps.
The
outer (inner) wheel is divided into seven (two) high voltage
sectors. In each of them a residual $\eta$-dependence of the response
will need to be corrected.


In the region devoted to precision physics, where the tracking
information is available ($1.5<|\eta|<2.5$), the depth segmentation is
similar to the barrel one. At large rapidity, there are only two
compartments in depth.  Twelve adjacent electrodes are summed for one
strip cell and 3 (4) for middle and back cells in the outer (inner)
wheel. The segmentation of the EMEC is summarized in
Table~\ref{tab:granularity}. There are 3888 readout cells per module.
HV boards distribute the high voltage to $\phi$-sectors of 24 (4)
electrodes in the outer (inner) wheel, independently on each side.

\begin{table*}[t]
\begin{center}
\vspace*{0.5cm}
\begin{tabular}{|l|l||c|c|c|}
\hline
Modules & $\eta$ range &       Front            &     Middle          & 
           Back                     \\
\hline
\hline
 & [0.0,0.8]\&\ [0.8,1.35]& \ \ $0.003\times 0.1$       & $0.025\times 0.025$  &  $0.050\times 0.025$ \\
Barrel & [1.35,1.4]& \ \ $0.003\times 0.1$       & $0.025\times 0.025$  & - \\
 & [1.4,1.475]& \ \ $0.025\times 0.1$       & $0.075\times 0.025$  & - \\
\hline
\hline
      & [1.375,1.425]& \ \ $0.050\times 0.1$       & $0.050\times 0.025$  & 
               --                   \\
      & [1.425,1.5]  & \ \ $0.025\times 0.1$      & $0.025\times 0.025$ & 
               --                   \\
EC & [1.5,1.8]    & $\sim 0.003\times 0.1$ & $0.025\times 0.025$ & 
               $0.050\times 0.025$  \\
Outer      & [1.8,2.0]    & $\sim 0.004\times 0.1$ & $0.025\times 0.025$ & 
               $0.050\times 0.025$  \\
      & [2.0,2.4]    & $\sim 0.006\times 0.1$ & $0.025\times 0.025$ & 
               $0.050\times 0.025$  \\
      & [2.4,2.5]    & \ \ $0.025\times 0.1$      & $0.025\times 0.025$ & 
               $0.050\times 0.025$  \\
\hline
EC Inner & [2.5,3.2]    & --                     & $0.1\times 0.1$     & 
               $0.1\times 0.1$      \\
\hline
\end{tabular}

\vspace*{0.5cm}
\caption{\it Granularity ($\Delta \eta \times \Delta \phi$) of
calorimeter cells in the electrodes A and B of the barrel modules and
the outer and inner wheels of the endcaps (EC). \label{tab:granularity}}
\end{center}
\end{table*}
\vspace*{0.5cm}

Three endcap modules where exposed to the H6 beam line namely ECC0,
ECC1 and ECC5.

\subsection{\bf Design modifications}

Following the construction and the beam tests of the prototype modules
0~\cite{NIMBARREL0,NIMENDCAP0}, a few design modifications have been
implemented :
\begin{itemize}

\item The high voltage is distributed on the electrodes by resistors
made of resistive paint silkscreened on the outer copper layer of the
electrodes. The role of these resistors is also to limit the current
flowing through the readout electronics in case of an unexpected
discharge of the calorimeter cell. Some of these resistors located
near the electrode bends were displaced in order to avoid breaking 
during the bending process. This lead to a change of the bending
process of the barrel electrodes~\cite{Electrodes}.

\item On the prototype, an increased cross talk and pulse
shape deformation was observed every 8 (4) channels in the barrel
(endcap). This effect was traced back to the absence of ground on one
side of some signal output electrode connectors. These missing
contacts were added in the production electrode circuits.

\item Summing and motherboards\footnote{Motherboards, plugged on top
of the summing boards, ensure the signal routing and the distribution
of calibration signals.} of the endcap were redesigned to reduce the
inductive cross talk to an acceptable level ($<$1~\%)~\cite{xtalkEC0}.

\item Precise calibration resistors (0.1~\% accuracy) are located on
the mother board. It appears that if an accidental HV discharge
occurs, a large fraction of the energy released could damage these
resistors, thus preventing any calibration of the cell. Diodes were
added on the signal path to protect these resistors for the middle and
back cells where the detector capacitances are the largest. For the
strips section such a protection scheme~\cite{Diode-Prot} could not be
installed, but the smaller detector capacitances reduce the risk of
damage.

\item Finally for the endcap construction, the cleanliness and
humidity regulation of the stacking room were improved. The honeycomb
spacers, which keep the electrodes centered on the gap, were modified
to ensure a better positioning. They were measured and tested under
high voltage before stacking and checked at the nominal high voltage
settings after each gap stacking.

\end{itemize}

\subsection{\bf Quality control measurements}

During the production a number of quality control measurements was
performed on all barrel modules. The lead thickness, the argon gap
width and some electrical parameters linked to the electrode design
are key parameters for the response of the calorimeter. These
measurements were done to ensure that all components were made with the
required precision.  Similar measurements were also performed on the
endcap modules, details of the gap variation will be discussed in the
endcap data analysis section.

\subsubsection{Absorber thickness}
\label{sec:leadthickness}
 
\begin{figure}[htbp]
\begin{center}
\includegraphics[width=10.cm]{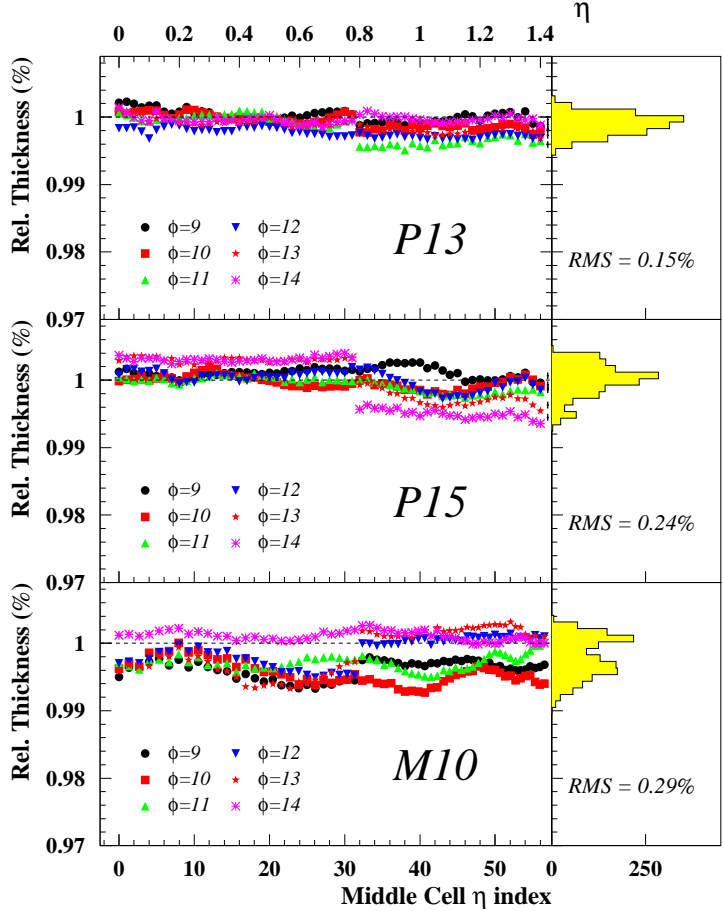}
\caption{\it Weighted average lead absorber thickness per middle cell
as a function of the middle cell $\eta$ index for various $\phi$
regions. The distribution of the measurements is also shown and the
dispersion is indicated.}
\label{fig:leadwidths}
\end{center}
\end{figure}

Before assembling the modules, the thickness of all lead sheets were
measured by means of an ultra sound system, producing a map with a 5
$\times$ 5 cm$^2$
granularity~\cite{Barrel-construction,absorbeursECAP}. These maps have
been used to select the stacking order of the lead sheets to minimize
the thickness fluctuations across each module. For each lead plate the
measurements are averaged projectively along the shower axis. As most
of the shower energy is shared by five absorbers, the thicknesses for
$|\eta| < 0.8$ ($|\eta| > 0.8$) normalized to their nominal values are
averaged in a sliding window of five consecutive absorbers. Finally to
emulate the energy profile within the electron cluster, each average
is weighted according to a typical electron energy deposition
shape. In Fig.~\ref{fig:leadwidths} the variation of the normalized
lead thickness as a function of the middle cell $\eta$ index (as
described below) for the modules P13, P15 and M10 is shown. The
dispersions do not exceed 0.3~\%\ per module.

For barrel modules the $\eta$ middle cell indices run from 0 to 54
covering the pseudo rapidity range from 0 to 1.35. The $\phi$ cell
indices correspond to increments in azimuthal angle of $\pi/128$. They
range between 0 and 15. For endcap modules the middle cell $\eta$
index used covers the region in pseudo rapidity from 1.375 up to 3.2
with indices ranging from 0 to 50. The $\phi$ middle cell indices
increment is the same as that of the barrel modules for the outer
wheel and $\pi/32$ for the inner wheel. In the $\phi$ direction these
indices range from 0 to 31 in the outer wheel and from 0 to 7 in the
inner wheel.



Similar measurements were also performed on endcap
modules. The dispersions of the normalized lead
thicknesses are within 0.3~\% for all modules.

\subsubsection{Gap dispersion}

At the end of modules stacking, the gap capacitance of each sector of
$\Delta \eta =0.2$, corresponding to eight middle layer cells, was
measured.  In each of these sectors, the capacitances were normalized
to their average value and the gap variations were extracted. Similarly
to the lead thickness, most of the shower is shared in a few
gaps. Using a typical electromagnetic lateral shower energy profile, a
sliding energy-weighted-gap was calculated. Its dispersion is
summarized in Table~\ref{capadisp} for the 3 barrel modules. While P13
and P15 show similar results, the module M10 has a larger
dispersion. This effect was explained by the use of electrodes made
before and after the modification of the bending process.


\begin{table}[htb]
\vspace*{0.5cm}
\begin{center}
\begin{tabular}{|c|c|c|c|}
\hline
Module & P13 & P15 & M10 \\ \hline \hline
Total & 0.62~\%\ & 0.64~\%\ & 1.66~\%\ \\ \hline
FT0 subset   & 0.58~\%\  & 0.39~\%\  & 1.41~\%\ \\ \hline 
\end{tabular}

\vspace*{.3cm}
\caption{\it Dispersions of the weighted average gaps for each barrel
module P13, P15 and M10. The FT0 subset corresponds to the region
instrumented with the readout electronics used in the analyses
described in Sec.~\ref{sec:readoutelec} and
Sec.~\ref{sec:modulesunif}.\label{capadisp}}
\end{center}
\end{table}
\vspace*{0.5cm}

Since in endcap modules the size of the gap is not constant, the
precise measurement of the gap is very important in order to evaluate
the electron clusters energies. The gaps were also indirectly
estimated from the cell capacitance measurements in the endcap
modules~\cite{mesureselecECAP}.  These results are discussed in
Sec.~\ref{sec:capacor}.

\section{CALIBRATION AND SIGNAL RECONSTRUCTION} 
\label{Sec:SignalRec}

\subsection{\bf Readout Electronics}
\label{sec:readoutelec}

As described in Sec~\ref{BarrelMod}, the readout signal consists of
the sum of signals measured in various electrodes. The analog sum is
made by summing boards connected at both ends of the electrodes; at
the front for the strip compartment and at the back for the middle and
back compartments. The motherboards, plugged on top of the summing
boards, send the signals through cold cables out to the front-end
boards placed outside the cryostat at room temperature. The readout
electronics are located in crates at both ends of the cryostat. To
extract the calorimeter signals from inside the cold vessel the
feedthrough (FT) connexion devices are used~\cite{TDR-LARG}. Two FTs
are required for each barrel module. Each of them covers half a
module in $\phi$. Three FTs are required for each endcap module. The
readout front-end electronics crates are placed on top of the FT
devices. In particular, for the beam test barrel modules there are two
FT devices covering a half barrel each. The regions covered are
denoted FT-1 and FT0.

In the front-end boards, signals are amplified and shaped through a
bipolar $\mathrm{CR\cdot RC^2}$ filter with a time constant $\tau_S =
15$~ns, then sampled every 25~ns and digitized. The shaping consists
in one derivation (CR) to form a bi-polar signal and two integrations
(RC) allowing both to reduce the impact of the pile-up and electronic
noise. The choice of shaping time constant results from the
minimization of the electronic noise and pile-up.

The energy dynamical range covered by the calorimeter signals requires
16~$bits$ whereas the available Analog-to-Digital Converter (ADC)
modules are limited to 12~$bits$. To produce
the adequate signal range the shapers produce three signals amplified
in three different gains low, medium and high in the ratio of
respectively 1, 9.3 and 93. All signals are then stored in an analog
pipeline awaiting for a first level trigger decision. When the
decision is made the signals either from the three gains or from the
most suited one according to a hardware gain selection are
digitized. In the beam test setup the signals are directly readout
while in the full ATLAS configuration they are sent to a higher level
firmware system where the energy is fully reconstructed for further
use both higher level trigger and in physics analysis.

\subsection{\bf Detector Electrical Properties} 
\label{Sec:DetElPr}

\begin{figure}[thbp]
\begin{center}
\includegraphics[width=9cm]{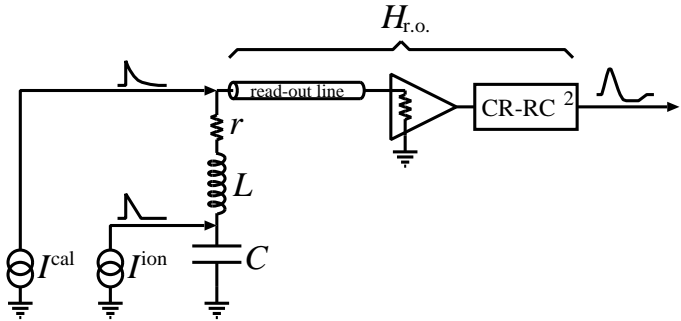}
\caption{\it Simple schematic electrical model of the cell electronic
environment. The shape of the calibration and ionization signals is
also illustrated along with the output pulse. Here $C$ is the cell
capacitance, $L$ the inductive path of the signal and $r$ is the
resistance.}
\label{fig:sigrec:Model}
\end{center}
\end{figure}

\begin{figure}[thbp]
\begin{center}
\includegraphics[width=11cm]{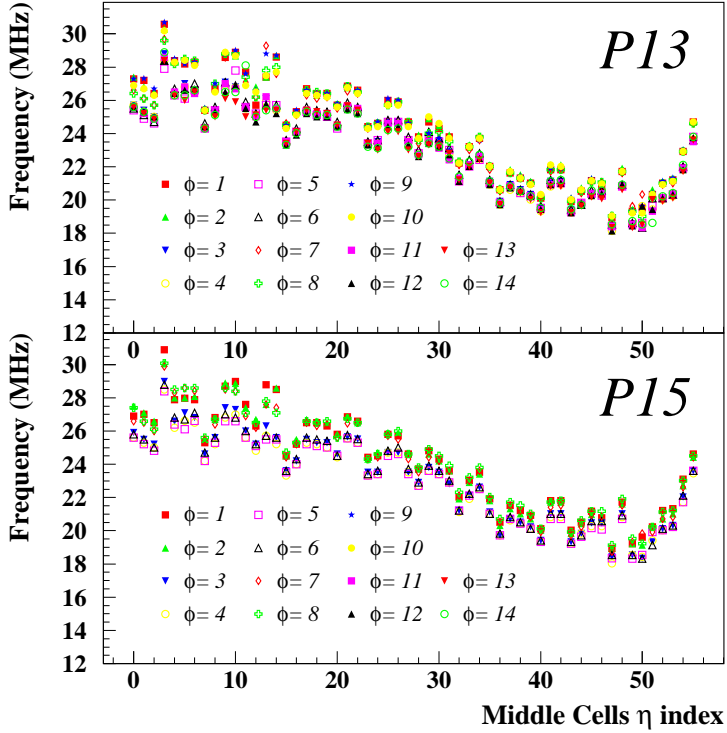}
\caption{\it Resonance frequencies for modules P13 and P15 as a
function of the $\eta$ index for all middle cells for various $\phi$
regions.}
\label{LC-nw}
\end{center}
\end{figure}

A detector cell can be seen as a resonant circuit as illustrated in
Fig.~\ref{fig:sigrec:Model}. The resonance frequency is linked to the
capacitance and inductance of the detector cell and summing
boards. These resonance frequencies have been measured on the
assembled calorimeter for each module using a network
analyzer~\cite{stephanie}. In Fig.~\ref{LC-nw} the results for the
module P13 and P15 filled with liquid argon are displayed. The
structure is mainly due to the variation of the inductance along
$\eta$ resulting from the electrode design in which the lines serving
to extract the signal of middle cells varies with $\eta$. The
differencies along $\phi$ reflect non uniformities of the summing
boards. The $\phi$ dispersion of these frequencies amounts to about
1.2~\%, which is compatible with the expected gap variation.

Similar measurements were made on endcap modules yielding frequencies
increasing as a function of $\eta$ from 20 to 40~MHz.

\subsection{\bf Calibration System}
\label{sec:CalibrationSystem}

The calorimeter is equipped with an electronic calibration
system~\cite{sigrec:ATL-LARG-2000-006} that allows prompt measures of
the gain and electrical response of each readout cell. This system is
based on the ability to inject into the detector an
exponential\footnote{A small offset $f_s$ is present, due to the
resistive component of the inductance in the calibration
board. Typically $f_s\simeq 7~\%$.}  calibration pulse of known
amplitude ($\Ical(t)$) that mimics the ionization pulse generated by
the particles hitting the detector:
\begin{equation}
  \Ical(t)=\Ical_0\theta(t)\left[f_s+(1-f_s)\Exp{-\frac{t}{\Tc}}\right]
  \label{eq:gcali}
\end{equation}
with a time decay constant $\Tc\simeq350$~ns ($\theta(t)$ is the unit
step function). The calibrated pulse shape ($\gcal(t)$) is
reconstructed using programmable delays by steps of about 1~ns.

The signal is generated by means of a digital-to-analog converter
(DAC) which controls the input current. Its value is proportional to
the DAC requested. A constant very small parasitic charge injection
DAC value is present due to parasitic couplings. It was
measured and its value accounted for.

The signal is produced in calibration boards placed in the front-end
crates. It is then carried into the cold vessel and distributed to the
electrodes through injection resistors $R_{inj}$ which are precise
at the 0.1~\%\ level. These resistors are placed on the motherboards.

A non uniformity of the calibration signals would affect the
uniformity {\it in fine}. All potential sources of non uniformity
affecting the calibration signal have thus been independently
measured. The main ones are:
\begin{itemize}
\item[(i)] the pulsers: each line is measured on the calibration
board directly. A relative dispersion of 0.19~\%\ is found.
\item[(ii)] calibration resistors: each calibration resistor is
measured and a relative dispersion of 0.08~\%\ is found for barrel
modules and 0.05~\% in the endcaps.
\item[(iii)] cables: The attenuation by skin effect which is
proportional to the cable length is corrected in average, however a
small relative dispersion of at most 0.1~\%\ could still impact the
energy uniformity.
\end{itemize}

The overall precision of the calibration system is $\sim$0.23~\%.

\subsection{\bf Energy Reconstruction}

When electrons and photons hit the calorimeter they interact within
the lead absorbers producing an electromagnetic cascade. Its charged
component ionizes the \LAr\ in the gaps, inducing a triangular current
signal~\cite{sigrec:Radeka74} ($\Iion(t)$):
\begin{equation}
  \Iion(t)=\Iion_0\theta(t)\theta(\Td-t)\left(1-\frac{t}{\Td}\right)
  \label{eq:gphys}
\end{equation}
whose length equals the drift time, $\Td\simeq450~\mathrm{ns}$ (in the
endcaps the drift time decreases from 550 down to 250~ns as a function
of $\eta$).

The ionization signal amplitude is reconstructed in each gain from the
five digitized samples $S_k$, to which the pedestals have been
subtracted (see Sec.~\ref{sec:pedestals}), located around the peak,
using the {\em optimal filtering} (OF)
technique~\cite{sigrec:Cleland84}:
\begin{equation}
  \left\{ \begin{array}{rcl}
   A     &=& \sum_{k=1}^{5} a_kS_k \\
   A \times \tau  &=& \sum_{k=1}^{5} b_kS_k \\
   \end{array}\right.
   \label{eq:sigrec:OF}
\end{equation}
where $A$ is the amplitude estimator and $\tau$ is the
signal arrival time estimate with respect to the readout clock. The
coefficients $a_k$ and $b_k$ are chosen in order to minimize the
effect of electronic noise.
They are analytically calculated through a Lagrange multiplier
technique \cite{sigrec:Cleland84}, provided one knows for each readout
cell the normalized shape of the ionization signal $\gion(t)$, its
derivative and the noise time autocorrelation. The latter is computed
from the data acquired during the pedestal runs.

The cell energy is then reconstructed from the cell signal amplitude
$A$ using the following prescription:

$$E_{vis}^{cell}=\frac{1}{f_{I/E}} \xi A $$

This formula can be read as the sequence of the following operations:
(i) the conversion of the signal in ADC counts $A$ into a current in
$\mu$A corresponding to the calibration of the readout electronics
with the function $\xi$; (ii) the conversion factor from current to
energy $f_{I/E}$.

\subsubsection{Pedestal Subtraction}
\label{sec:pedestals}

During the running period, data in absence of beam were taken daily to
assess the level of signal without any energy deposition in each cell
for all gains of all modules. In addition, in order to address
precisely the more subtle possible variations on a run-by-run basis
the pedestals can be evaluated from triggers taken at random during
the run. The variations between no-beam runs and random triggers are
in general negligible and thus not taken into account but for a few
runs small instabilities, in particular in the presampler, could
produce a detectable bias in energy not exceeding 20~MeV. The
subtraction of pedestals is done on all signal samples ($S_k$) before
applying the {\em optimal filtering} method.

\subsubsection{Calibration Procedure}

The calibration procedure establishes the correspondence between a
signal readout in ADC counts and a known injected current in the cell
in $\mu$A. Calibration data were taken approximately twice daily
throughout the running period.

The procedure consists in fitting the ADC response as a function of
the DAC values knowing that the injection currents vary linearly with
the DAC values. A second order polynomial form $\xi$ is used. The
higher order fine structure of the non linear response of the
calibration is not relevant here. The aforementioned function also
contains a DAC to current constant conversion factor.

\subsubsection{Current to Energy conversion Factor}
\label{sec:muA2GeV}

Estimating from first principles the relation between the measured
current and the energy is an intricate task, as numerous complex
effects can introduce biases as detailed in~\cite{linearity}. However,
the simplified model estimation of $f_{I/E} \sim 14.4~$nA$/$MeV in the
barrel accordion in the straight sections (7~\% less when taking into
account the folds~\cite{FoldsEffect}), yields a result accurate to the few percent
level. This value is in agreement with a more precise calculation in
which the detailed cell electric fields and recombination effects are
taken into account. In the case of the analysis of the barrel modules
where a complete and thorough simulation of the calorimeter was used a
more precise estimate of the $f_{I/E}$ factor is obtained from a
comparison of the Monte Carlo simulation with the data (see
Sec.~\ref{sec:dataMCcomp}).

\subsection{\bf Signal Reconstruction}              
\label{sec:signalreco}

\begin{figure}[thbp]
\begin{center}
\includegraphics[width=9cm]{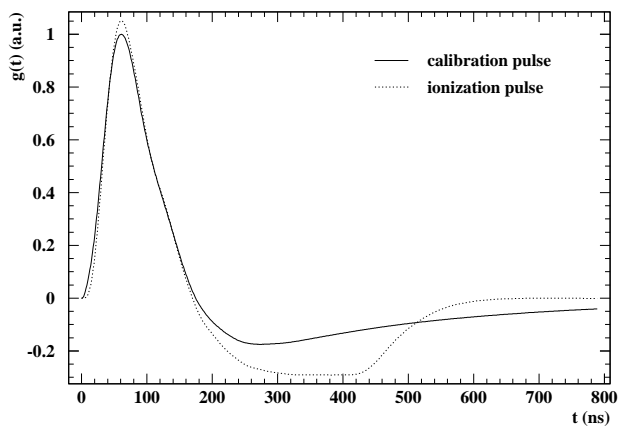}
\caption{\it Normalized calibration $\gion(t)$ and ionization $\gcal(t)$ pulses.}
\label{fig:sigrec:Shapes}
\end{center}
\end{figure}

The normalized response $\gion(t)$ of the system to the ionization
current differs from the response $\gcal(t)$ to a calibration
current because the two pulses are respectively triangular and
exponential, and while the former is generated inside the detector,
the latter is injected in the cell from one end of the detector and
reaches the inside through an inductive path\footnote{The effect is
not negligible, as it affects the amplitude ratio of the ionization
and calibration waveforms by approximately 0.15~\%/nH, and the
inductance value varies throughout the detector from 35 to
55~nH~\cite{sigrec:ATL-LARG-2001-018}.}. Typical shapes of the two
signals at the end of the readout chain are shown in
Fig.~\ref{fig:sigrec:Shapes}.

The difference between the calibration and the ionization pulses can
be analytically described by modeling the \LAr\ readout cell as a
lumped $rLC$ model (see Fig.~\ref{fig:sigrec:Model}), where the two
currents share the same readout chain while being generated in
different places. After taking into account the analytical
descriptions (\ref{eq:gphys}) and (\ref{eq:gcali}), the relation
between the current shapes is
\cite{sigrec:ATL-LARG-2001-008,sigrec:ATL-LARG-2004-007,sigrec:ATL-LARG-2005-001}:
%
%
\begin{eqnarray}
  \gion(t) = \gcal(t) &\times& \invLapl{\frac{s\Td-1+\mathrm{e}^{-s\Td}}{s^2\Td} 
	                  \frac{s(1+s\Tc)}{f_s+s\Tc}} \nonumber \\
                      &\times& \invLapl{\frac{1}{1+s^2LC+srC}} 
  \label{eq:sigrec:gphys_pred_time}
\end{eqnarray}

where the normalized ionization signal $\gion(t)$ can be inferred from
the observed calibration signal $\gcal(t)$ by means of time-domain
convolutions with functions that depend on the parameters
$\Td,\Tc,f_s,LC,rC$ (where $\Tc$, $f_s$ are the calibration time
constant and a calibration offset) only using the inverse Laplace
transform $\mathcal{L}^{-1}$ with the Laplace variable $s$. The
resonance frequencies displayed in Fig.~\ref{LC-nw} correspond to the
standard oscillator circuit thus $f=1/2\pi\sqrt{LC}$. The evaluation
of $\gion(t)$ is completely independent of any details of the read-out
chain.

For the barrel modules the drift time $\Td$ can be estimated with
various
methods~\cite{sigrec:ATL-LARG-2004-007,sigrec:ATL-LARG-1995-029,sigrec:ATL-LARG-99-009},
while the parameters $\Tc, f_s$ and $LC,rC$ can be extracted either by
analyzing the observed signals or from direct measurements. In the
endcaps $\Td$ varies continuously with $\eta$, it was thus considered
as a parameter. Two strategies were developed for the test beam, as
described in the next subsections.

\subsubsection{Semi-predictive approach} \label{sec:sigrec:TCM}

The following method
\cite{sigrec:ATL-LARG-2001-008,sigrec:ATL-LARG-2005-001} has been used
to reconstruct the energy for the prototype
modules~\cite{NIMBARREL0,NIMENDCAP0} and for all tested production
modules.

At the test beam, the electrons reach the calorimeter at random time
with respect to the sampling clock\footnote{ The time between the
trigger given by a scintillator coincidence and the 40 MHz clock edge
is measured for each event.} (asynchronous events).  It is thus
possible to sample the ionization signal every nanosecond, similarly
to what is done for the calibration signal. However, the signal shape
obtained from direct observation is imprecise, due to the low
statistics and the large fluctuations in the shower
development. Moreover, its normalization is arbitrary.

The parameters $\Tc$ and $f$ are measured directly on the calibration board
before its installation on the detector.

The parameters $LC$ and $rC$ are obtained from a fit of the predicted
ionization signal, as obtained from
equation~(\ref{eq:sigrec:gphys_pred_time}), to the observed one. In
the fit, two more parameters are allowed to vary in order, to account
for a time shift between the two signals and for the amplitude scale
factor. Once $LC$ and $rC$ are found, the predicted ionization signal
has the correct normalization and can be used to evaluate the OF
coefficients.

\subsubsection{Fully predictive method}  \label{sec:sigrec:RTM}

This method is an alternative to the one described previously: it has the
advantage of being based on calibration data only, thus not relying on a direct
knowledge of ionization pulses from asynchronous events. It relies on the
observation of long enough calibration signals: up to 32 digitized samples can
be acquired, corresponding to a maximum length of 800~ns. The details are fully
described in~\cite{sigrec:ATL-LARG-2004-007}, therefore only an overview is
given here.

The exponential decay time can be extracted from a fit of the tail of
the calibration signal. The offset $f$ can be estimated as follows: if
the injected $\Ical(t)$ was a step-function, then the tail of the
shaped signal would be minimal. The detector response to an injected
step could be calculated, by means of a time-domain convolution
between $\gcal(t)$ and a function of time which depends on the
parameters $\Tc$ and $f$. Once $\Tc$ is found, the best value for $f$ is
chosen as that minimizing the tail of the response function.

The parameters $LC$ and $rC$ are extracted from a frequency analysis
of the transfer function, which exhibits a minimum for angular
frequency $\omega=\frac{1}{\sqrt{LC}}$. This is achieved either by a
direct use of a fast Fourier transform, or with techniques similar to
that described for the extraction of $f$ --- here the characteristics of
the detector response to a sinusoidal injected signal are exploited:
the best value for $\omega$ is that minimizing the oscillations in the
tail.

Such a technique has been applied to a restricted region of a
production module where the 32-samples-calibration data were made
available (half a module for the Middle and Back compartments, and
only a $\Delta\eta\times\Delta\phi=0.2\times0.2$ sector for the
Strips). The residuals between the observed and predicted ionization
signal are at the level of 1~\% (0.2~\% in the peak region). The
agreement in the timing between the OF reconstruction and the
scintillator measurement is within 350~ps.


For various runs taken on a reduced region of the barrel modules both
methods were applied. The comparison of the two methods is used to
estimate the possible biases due to the reconstruction method
chosen. The results of this comparison are given in
Sec.~\ref{sec:enerecoscheme}.

\subsubsection{Prediction of the Physics to Calibration Amplitude Ratio}  
\label{sec:MphysMcal}

\begin{figure}[thbp]
\begin{center}
\includegraphics[width=11cm]{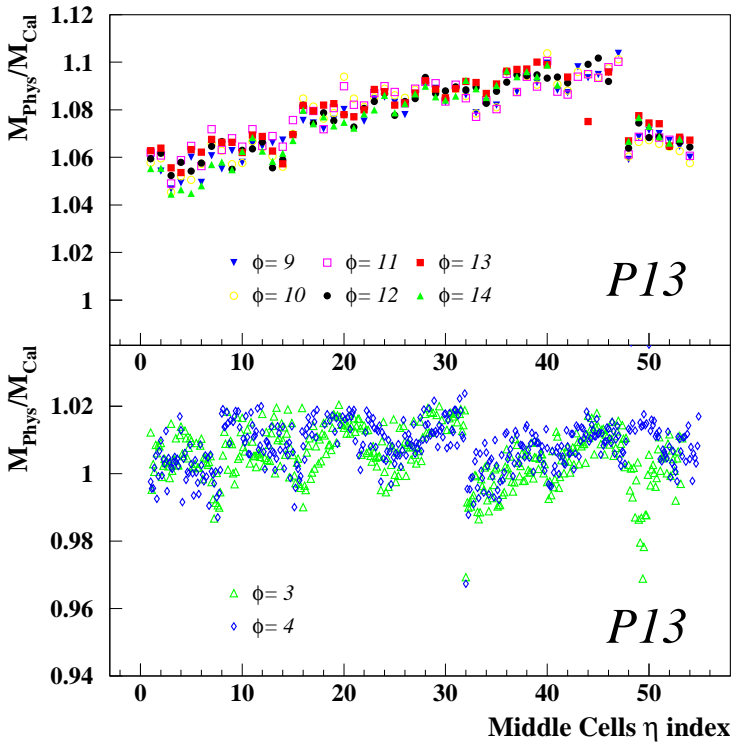}
\caption{\it Bias in the signal reconstruction method as derived for
the barrel P13 module at all middle $\eta$ indices and azimuthal
angles, for middle cells (upper plot) and strips (lower plot).}
\label{fig:MphysMcalBarrel}
\end{center}
\end{figure}

\begin{figure}[thbp]
\begin{center}
\includegraphics[width=11cm]{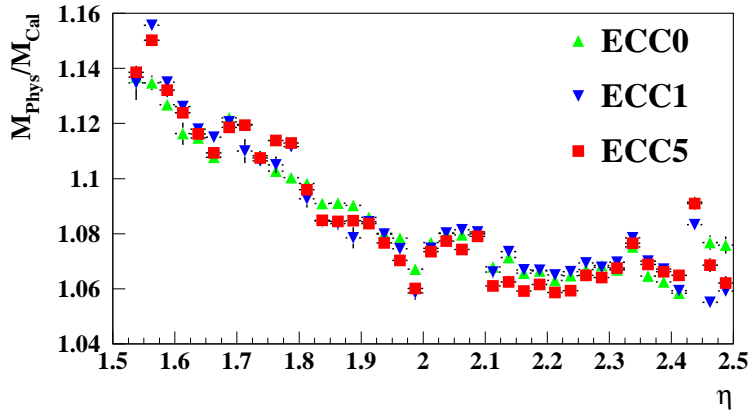}
\caption{\it Bias in the signal reconstruction as derived for all
tested EMEC modules (ECC0, ECC1 and ECC5) for middle cells averaged in
the azimuthal direction as a function of the pseudo rapidity.}
\label{fig:MphysMcalECAP}
\end{center}
\end{figure}

Since the predicted signal and calibration shapes are different, the
response amplitude to a normalized input signal will be
different. This difference must be taken into account in order to
correctly convert ADC counts into energies. It is done using the
prediction of the physics to calibration amplitude ratio, namely
$M_{phys}/M_{cal}$. This ratio varies with pseudo rapidity. These
variations are displayed in Fig.~\ref{fig:MphysMcalBarrel}
and~\ref{fig:MphysMcalECAP} for the barrel and endcap modules
respectively.

The barrel strips $M_{phys}/M_{cal}$ is reasonably consistent with 1,
whereas for middle cells the prediction to calibration amplitude ratio
increases systematically with $\eta$ up to the middle cell $\eta$
index 48. These variations are consistent with the resonance
frequencies measured and presented in Sec.~\ref{Sec:DetElPr}. A
similar effect is observed in endcap modules.


\subsubsection{Energy Dependence with Time}

Contrary to what will be the case in the ATLAS experiment and due to
the time spread of electron bunches, in the test beam the time phase
is essentially random. For all barrel and endcap modules it was
checked that in the energy reconstruction scheme no bias is observed
as a function of the time phase.

\subsection{\bf Cross Talk Issues}
\label{sec:crosstalks}

\subsubsection{Module Electrodes Cross Talks}

Various cross talk effects inherent to the design of calorimeter cells
or due to the readout electronics are present in the calorimeter. A
complete description of the origin of these effects can be found
in~\cite{JColas,xtalkEC0}. These unavoidable effects have been measured
mainly using calibration
signals~\cite{sigrec:ATL-LARG-2000-007,xtalkBarrel,xtalkEC}. A summary
of the measurements of the typical cross talks are given in
Table~\ref{Barrelcrosstalktable} for both the barrel and the endcap
electrodes. The cross talks are here defined as peak-to-peak, {\it
i.e.}, the maximum amplitude of the cross talk is normalized to the
signal amplitude. All the effects described here are those measured in
nearest neighbors, second order effects are negligible, except for the
secondary cross talk between a strip and its next to nearest neighbor
with a peak-to-peak value of 0.9~\% for barrel and 0.5~\%\ for endcap
electrodes.

Among these cross talk effects mainly two will affect the energy
reconstruction: the strips capacitive cross talk and the middle-back
inductive cross talk.

The strips cross talk effect is due to the thin separation between the
finely segmented cells of the first compartment of the calorimeter,
the fine segmentation is necessary for the $\pi^0/\gamma$ separation,
for a precise estimation of the pseudo rapidity of the impact point
and for the estimation of the photon pointing direction. This effect
non trivially affects the calibration and the signal
reconstruction. Its treatment is described in
Section~\ref{StripsCrossTalk}.

The cross talk between the middle and the back compartment results
from a mixture of various effects, but is mostly inductive. Since the
back compartment plays an important role in the assessment of the
longitudinal shower energy leakage it is important that this effect is
measured and corrected for.

All these effects were expected and are well understood. They were
measured on all modules. All cross talk effects are well reproducible
among modules.

\begin{table}[tbp]
\begin{center}
\vspace*{.3cm}
\begin{tabular}{|c|c|c|c|c|c|c|}
\hline
Module Part & \multicolumn{3}{c|}{Electrode A} &  \multicolumn{3}{c|}{Electrode B} \\
\hline
Compartment & Front & Middle & Back & Front & Middle & Back \\
\hline
Front       & 6.9~\%\  $^2$ & 0.07~\%\ $^1$ &  0.04~\%\ $^4$  & 6.9~\%\ $^2$  & 0.09~\%\ $^1$  & 0.04~\%\ $^4$ \\
Middle      & 0.07~\%\ $^1$  & 1.5~\%\ $^{2+3}$ &  0.5~\%\ $^{3+2}$  & 0.09~\%\ $^1$  & 1.5~\%\ $^{2+3}$  & 0.7~\%\ $^{3+2}$  \\
Back        & 0.04~\%\ $^4$  & 0.5~\%\ $^{3+2}$ &  1.9~\%\ $^3$  & 0.04~\%\ $^4$  & 0.7~\%\ $^{3+2}$  & 1.9~\%\ $^3$  \\
\hline
\hline
Module Part & \multicolumn{3}{c|}{Outer Wheel} &  \multicolumn{3}{c|}{Inner Wheel} \\
\hline
Compartment & Front & Middle & Back & Front & Middle & Back \\ 
\hline
Front       & 5-8~\%\ $^2$       & 0.2~\%\ $^1$ & 0.01~\%\ $^4$ & - & -      & - \\
Middle      & 0.2~\%\ $^1$       & 1.0~\%\ $^{2+3}$ & 1.0~\%\ $^{3+2}$       & - & 1.0~\%\ $^{2+3}$   & 0.5~\%\ $^{3+2}$ \\
Back        & 0.2~\%\ $^4$ & 0.5~\%\ $^{3+2}$ & 3.0~\%\ $^3$         & - & 0.8~\%\ $^{3+2}$ & 2.0~\%\ $^3$ \\
\hline
\end{tabular}

\vspace*{0.5cm}
\caption{\label{Barrelcrosstalktable} \it Summary of the typical
cross talks measured in the different samplings of the barrel
electrodes A and B and the outer and inner wheel of the endcap. The
indices denote the nature of the cross talk where 1, 2, 3 and 4
correspond to resistive, capacitive, inductive and mixed
respectively.}
\end{center}
\end{table}
\vspace*{0.5cm}


\subsubsection{Treatment of Strips Capacitive Cross talk}
\label{StripsCrossTalk}

Because an electron cluster contains a large number of strips (see
Sec.~\ref{sec:clustering} and Sec.~\ref{sec:ECclustering}) almost all
the electron signal is contained in the cluster cells. Therefore any
signal exported from one strip to its neighboring strips is recovered
in the reconstructed cluster energy. The energy of an electron or
photon cluster is therefore at first order not sensitive to the cross
talk effect. However, the readout electronics are calibrated using
pulse patterns where one strip cell is pulsed out of four. To recover
the signal loss in the neighboring cells, the signal readout in the
two first neighbor unpulsed cells is added to the pulsed cell and the
average of the two next to nearest neighbor unpulsed cells are also
added to the pulsed cell. In doing so using calibration runs where the
signal is sampled in 32 time intervals of one nanosecond the shape of
a signal in absence of cross talk is emulated. This new shape is
used to derive the OF coefficients and comparing it to the signal
shape without applying the summing procedure a correction factor of
the ramp gains is derived. 
The amplitude of the correction is around 7~\%. The method
has proven to be linear as the correction factors appear to be
independent of the DAC signal applied.  When the complete correction
is applied, the $M_{Phys}/M_{Cal}$ factor recovers a value of
approximately 1 (its expected value if the signal reconstruction is
sound) as illustrated in Fig.~\ref{fig:MphysMcalBarrel}.

It was also checked that the method is not sensitive to gain
differences between neighboring cells. To avoid the possible
calibration differences between strip cells, the signal read in the
non-pulsed cells is first calibrated and then added. However, no
noticeable difference is seen between the two approaches.


Strip cells cross talk corrections show collective variations of
approximately 1~\%\ of the strips energy. The devised correction thus
improves the uniformity of the energy measurement for the total
electron energy of a few percent relative but has almost no impact on
the local energy resolution.

\subsubsection{Feedthrough Resistive Cross Talk}


For all modules tested in the beamline the same two FT devices
equipping the test beam cryostat were used. One unexpected and more
tedious cross talk effect appeared in the bottom barrel modules FT
which was lacking gold plated contacts. It exhibited an increase with
time of the ground resistance common to all channels inside the
connectors (64 channels), therefore a long range resistive cross-talk
appeared. This problem non trivially affected most of the channels
corresponding to that FT.


A correction procedure was derived which brought the uniformity of the
response in channels readout with this feedthrough close to that of
the other channels. However the data taken with the corrupted FT are
not considered in the final analyses. These data correspond to
electrons impinging at azimuthal angles $\phi$ smaller than 9 in
middle cell index units. To prevent this problem to appear in ATLAS,
all connectors were gold plated and the connections were verified with
the final detector.


\section{BEAM TESTS EXPERIMENTAL PROTOCOLS} 
\label{Sec:TBsetup}

\subsection{\bf Beamline Setups}

The barrel and endcap production modules have been exposed to beams at
CERN with 245~GeV electrons in the H8 line and 120~GeV electrons in
the H6 line respectively. The pion contamination of the electron beam
was discussed in~\cite{linearity}, it is far less relevant in this
context as a small bias would equally affect all the data. Similar
pion contamination rejection cuts as those used in~\cite{linearity}
were nevertheless used. Each experimental setup has been already
described in details in~\cite{NIMBARREL0} and~\cite{NIMENDCAP0}. Each
beam line is instrumented with multi wire proportional chambers to
extrapolate the particle impact on the calorimeter. The trigger is
built from the coincidence of three scintillators on the beam line
defining a maximal beam spot area of 4 cm$^2$. While for the barrel
the projectivity of the beam is ensured in both directions by a proper
movement of the cryostat, for the endcap the cryostat is moved along
$\eta$ but for the $\phi$ rotation the calorimeter is moved inside the
argon.  In the endcap the amount of dead material upstream of the
calorimeter is almost constant at all electron impact points and
amounts to 1.5~$X_0$. Thus it does not introduce any effect in the response
uniformity of the calorimeter. In the barrel, however, the amount of
dead material upstream continuously increases with $\eta$.

\subsection{\bf Scanning Procedure}

Runs are taken with different positions of the moving table in such
way that the complete module is exposed to the beam. The positions are
chosen in order to center projectively each middle cell of the modules
into the beam. For this reason the cell coverage is not completely
uniform. In particular, less electrons are impinging on the edge of
the cells than in the center. Results are therefore commonly presented
in units of middle cells ($\eta$,$\phi$) indices. Exceptions are made
for certain regions of the barrel calorimeter that have been scanned
with a half middle cell granularity.





\subsection{\bf Temperature Stability}

Argon temperatures were readout and recorded on the barrel setup: the
temperature stability is better than 10 mK, but the absolute temperature
differs from one module to another. A -2.0~\% correction of the mean
energy per degree is taken into account in the
analysis~\cite{sigrec:ATL-LARG-1995-029}.

\subsection{\bf Data Sets} 


For all barrel modules runs of 10\,000 events were recorded at each
cell position.

Additional runs were taken in order to perform systematic studies such
as cell-to-cell transitions, the electrode A to electrode B
transition, and to study the impact of changing front-end and
calibration boards in the front-end crates.


In the outer wheel of the EC modules only the region $1.525<|\eta|<2.4$
and $0.075<\phi<0.725$ (corresponding to
$6\leq\eta_{\mathrm{cell}}\leq 40$ and
$4\leq\phi_{\mathrm{cell}}\leq29$) was covered. In the inner wheel the
domain $2.6<|\eta|<3.1$ and $0.2<\phi<0.7$ (corresponding to
$45\leq\eta_{\mathrm{cell}}\leq49$ and
$2\leq\phi_{\mathrm{cell}}\leq6$) was covered representing 25 cells.



In the ECC0 and ECC1 modules high voltages were incorrectly cabled in
the inner wheel. The inner wheel uniformity was therefore only studied
on ECC5.






In all modules a few single isolated channels were defectuous or not
responding at all. Most of them were due to the readout electronics
setup and were thus found in all running period for EC and barrel
modules. For those problematic channels for which the problem is
intrinsic to the modules, the modules have been repaired for their
future use in ATLAS.  For instance in the EC outer wheel, four (one)
electrode front connectors in ECC0 (ECC5) were not properly plugged on
their summing board. The corresponding cells ($\sim20$) are excluded
from the analysis. To avoid such problems in ATLAS, connections of all
production modules have been checked by specific measurements.

\begin{table}[htbp]
\vspace*{0.5cm}
\begin{center}
\begin{tabular}{|c|c|c|c|c|c|c|}
\hline 
Type & \multicolumn{3}{c|}{Barrel} & \multicolumn{3}{c|}{endcap} \\
\hline
    Set-up & M10 & P13 & P15 & ECC0 & ECC1 & ECC5 \\
\hline \hline
Bad Strips & 0 & 3 & 1 & 5 & 8 & 6 \\
\hline 
Bad Middle & 2 & 0 & 0 & 3 & 1 & 1 \\
\hline
Bad Calibration & 2 & 2 & 0 & \multicolumn{3}{c|}{0} \\
\hline
No data  & 2 & 1 & 15 & - & - & -\\
\hline
\end{tabular}

\vspace*{0.5cm}
\caption{\it Number of channels which present no physical signal,
those where no data were taken and those where the calibration line was
defectuous for all barrel modules in the FT0 region only and all
end-cap modules.}
\end{center}
\label{tab:dead}
\end{table}
\vspace*{0.5cm}



A problem which may appear in the future running of ATLAS is the
impossibility to run some sectors at the nominal values of high
voltage (HV). In the EC modules the electrodes of three HV sectors
were powered on one side only because of HV problems that appeared at
cold. In this case the energy of the corresponding cells is simply
multiplied by a factor of $\sim$2. The resulting energy resolution is
degraded by $\sim20$~\% in these sectors and by $\sim40$~\% at the
\-$\phi$-transition with a good sector. The impact on the response
uniformity is negligible. The correction could be refined at the
\-$\phi$-transition with a good sector. These cells are kept in the
encap modules analysis.

For the barrel and endcap modules the scanned cells that are
neighboring a middle defectuous channel or which are behind a bad
strip are excluded from the analysis. The same procedure is applied
for problems in the calibration. When a back or a presampler cell is
defectuous no specific treatment is applied.

The regions covered in endcap and barrel modules are detailed in
Table~\ref{tab:region}. For the cells removed from the analysis, in
the future a special treatment involving an energy correction could be
applied.


\begin{table}[tbp]
\vspace*{0.5cm}
\begin{center}
\begin{tabular}{|c||c|c|c||c|c|c|}
\hline 
    & \multicolumn{3}{c||}{Barrel}& \multicolumn{3}{c|}{EC Outer Wheel (IW)}\\
\hline 
 Module  & M10 & P13 & P15 & ECC0 & ECC1 & ECC5 \\
\hline 
Scanned  & 324 & 324 & 324 & 874 & 910 & 844 (25) \\
\hline 
   Kept  & 278 & 305 & 299 & 799 & 840 & 816 (25) \\
         & 86~\% & 94~\% & 92~\% & 91~\% & 92~\% & 97~\% (100~\%) \\
\hline
\end{tabular}

\vspace*{0.5cm}
\caption{\it Number of scanned cells that are kept for the uniformity
analysis.  The excluded cells are related to problems that are
specific to the beam test. The numbers indicated in parentheses
correspond to cells of the inner wheel. In the barrel section the
altogether 756 cells were scanned on each barrel modules comprising of
the FT0 and FT-1 but only those pertaining to the FT0 are
kept. \label{tab:region}}
\end{center}
\end{table}
\vspace*{0.5cm}

\section{BARREL UNIFORMITY} 
\label{Sec:BarrelUnif}

\subsection{\bf Monte Carlo Simulation} 
\label{Sec:MC}

As was done in~\cite{linearity} a full Monte Carlo description of the
shower development of electrons penetrating the electromagnetic
calorimeter barrel module has been carried out with the GEANT
simulation version 4.8~\cite{MC}. The intricate geometry of
the accordion and the material of the calorimeter is thoroughly
described. All the particles are followed in detail up to an
interaction range of 20$\mu$m. The photon-hadron interactions are
also simulated.

The material within the volume of the cryostat is thoroughly simulated
(lead, liquid argon, foam, cables, motherboards and G10) using our
best knowledge of its geometrical distribution as described
in~\cite{TDR-LARG}.

The absorber thickness and gaps were set to their nominal values as
reported in~\cite{NIMBARREL0} and not to the measured ones. The
material contraction in cold liquid argon was not taken into account
either. The main consequence of these small inaccuracies essentially
results in a absolute scale effect of a few per mil and does not
affect the uniformity.



The material outside and in front of the calorimeter is described in
detail accounting for the energy lost near and far from the
calorimeter. The case where bremsstrahlung photons are not
reconstructed in the calorimeter, because they have been produced far
upstream of the impact point, is thus taken into account.

\begin{figure}[htbp]
\begin{center}
\includegraphics[width=12.5cm]{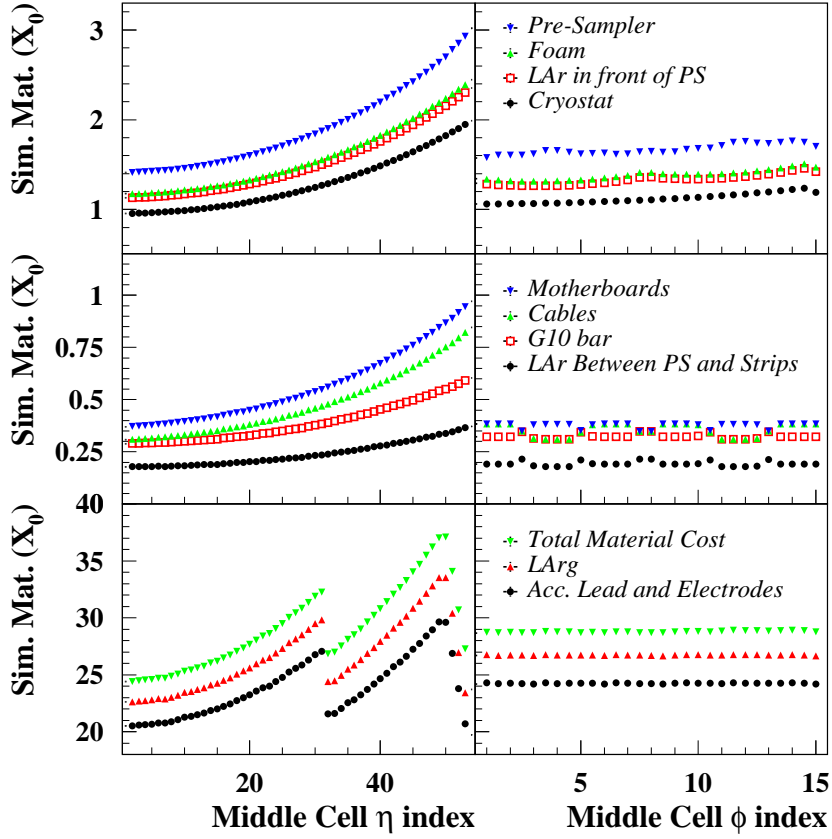} 
\caption{\it Distributions of the simulated material for the main
material types in units of radiation lengths ($X_0$) as a function of
$\eta$ and $\phi$. The upper two plots represent the cumulative
amounts of material before and in the presampler. The two intermediate
plots represent the cumulative distributions of the material between
the presampler and the strips compartment of the accordion
calorimeter. In order to illustrate the various structures in $\phi$,
in the $\eta$ plots the average values correspond to various fixed
$\phi$ values, the corresponding averages are thus different in
$\eta$ and $\phi$. The bottom two plots represent the amount of
material in the accordion calorimeter as well as the total amount of
material in the entire setup.}
\label{fig:materialcost}
\end{center}
\end{figure}

The material distribution of the test beam set-up is illustrated in
Fig.~\ref{fig:materialcost}. The distributions are presented as a
function of the $\eta$ and $\phi$ direction separately and
cumulatively for the material before and in the presampler, the
material between the presampler and the first accordion compartment
and the material in the accordion calorimeter. These distributions
correspond to Monte Carlo samples simulated along the $\eta$ and
$\phi$ directions with a granularity of one middle cell. Each point
contains at least five hundred events. To further illustrate the
material along the $\phi$ direction scans at various fixed $\eta$
values are performed. It appears that the total amount of material is
rather uniform in the $\phi$ direction.

Unlike in the ATLAS experiment, in the test beam setup only a small
amount of material is located upstream of the calorimeter. It is
therefore a unique opportunity to test the simulation of the
calorimeter alone.

In particular, the material upstream has been tuned by varying the
amount of liquid argon in front of the presampler that is not well
known from the construction. This allowed to optimize the agreement of
the Monte Carlo with the data in the presampler alone. As was the case
in~\cite{linearity} the best agreement is found for a thickness of
2~cm.

The large capacitive cross talk between strips is not simulated. The
cross talk effect between the middle and back compartments of the
calorimeter is taken into account in the simulation.

\subsection{\bf Clustering}
\label{sec:clustering}

The electron energy is reconstructed by summing the calibrated cell
energies deposited in the three calorimeter compartments and in the
presampler. A cluster is built around the cell with the largest energy
deposit in the middle compartment. The cluster size, expressed in
number of cells in $\Delta \eta_{cell} \times \Delta \phi_{cell}$ is
$3\times3$ throughout the barrel. This choice reflects a good
compromise between noise and energy containment.  The energies
deposited in the corresponding $2\times3$ back cells located behind
the middle cells of the cluster are then added. In the front
compartment of the barrel all 24 corresponding strip cells in front of
the middle cluster are added. In $\phi$, the cluster contains 1 or 2
cells, depending on whether the shower develops near or far from the
cell center. In the presampler the corresponding 3 cells in $\eta$
within the two corresponding $\phi$ positions are chosen to be part of
the cluster.

\subsection{\bf Energy Reconstruction Scheme}
\label{sec:eneschemeMC}

Several electron energy reconstruction schemes were tried. In
particular the two most effective ones were those used
in~\cite{linearity} and an additional reconstruction scheme which took
into account the shower depth dependence of the sampling fractions and
the leakage energy~\cite{giacomo}. The energy reconstruction scheme
used in~\cite{linearity} was chosen for its simplicity as it could be
applied across the entire $\eta$ range with a simple analytical
parametrization. This energy reconstruction scheme is mostly based on
our best knowledge of the detector as implemented in the
Monte Carlo simulation. The total reconstructed
electromagnetic (EM) shower energy $E_{rec}$ is evaluated from the
measurements of the visible cell energies in the presampler
($E_{PS}^{meas}=\sum_{PS} E_{vis}^{cell}$), all compartments of the
accordion added together ($E_{Acc}^{meas}=\sum_{Acc} E_{vis}^{cell}$), the
energy measured in the strips ($E_{Strips}^{meas}=\sum_{Strips}
E_{vis}^{cell}$) and the energy measured in the back compartment
($E_{Back}^{meas}=\sum_{Back} E_{vis}^{cell}$). These measurements are
carried out within the EM cluster.

The basic principles of the energy scheme for the reconstruction of
test beam electrons are reviewed in~\cite{linearity}. The total
deposited energy is reconstructed in four steps: (1) the energy
upstream of the presampler is evaluated using the measured presampler
energy; (2) the energy deposited between the presampler and the strips
is evaluated using the measured presampler and strips energies; (3)
the energy deposited in the accordion is evaluated using the measured
energy in the accordion cells; (4) the leakage energy is evaluated
from the average expected leakage at a given position in the detector
and the amount of energy in the last accordion compartment. The scheme
can be written as follows:

\begin{equation} E^{rec}_{raw}  =  a_{\eta}+b_{\eta}\times E_{PS}^{meas} +  
c_{\eta} \times \sqrt{E_{PS}^{meas} E_{Strips}^{meas}} + 
\frac{E_{Acc}^{meas}}{d_{eta}}
+\xi(E_{Back}^{meas})
\label{fullparam}
\end{equation}

\noindent yielding the raw reconstructed energy, corresponding to the
complete raw shower energy. 


As shown in Section~\ref{Sec:MC}, the amount of material is
rather uniform in the azimuthal direction, therefore all parameters
are derived only as a function of the $\eta$ direction where the
variations of material are large.

As explained in detail in~\cite{linearity}, this energy scheme has
numerous advantages with respect to the one used
before~\cite{NIMBARREL0}. The most prominent are:

\begin{itemize}
\item[(i)] It optimizes both resolution and linearity.
\item[(ii)] The parametrization of the energy deposited between the
presampler and the strips compartment of the accordion allows to
sample a different part of the shower and to absorb most of the shower
depth dependence of the overall sampling fraction.
\item[(iii)] The offset in the parametrization of the energy deposited
before the presampler allows to account optimally for the energy loss
by ionization by the beam electrons.
\end{itemize}

The main differences with the energy scheme used in~\cite{linearity}
are the following:

\begin{itemize}
\item[(i)] The parameters are derived as a function of $\eta$ and
not as a function of the energy as the scan is done at fixed energy.
\item[(ii)] The leakage energy is derived from the energy in the
back compartment of the accordion. The electron energies in the
present analysis are higher than those used in~\cite{linearity} where
beam test runs were taken at a fixed $\eta$ value of 0.687
corresponding to a region of the detector where the longitudinal
leakage is the close to smallest. The expected leakage is much larger
here. It has to be thoroughly corrected in order to reach a good
energy resolution.
\end{itemize}

\subsubsection{Parametrization of the Calibration Parameters}

\begin{figure}[htbp]
\begin{center}
\includegraphics[width=9cm]{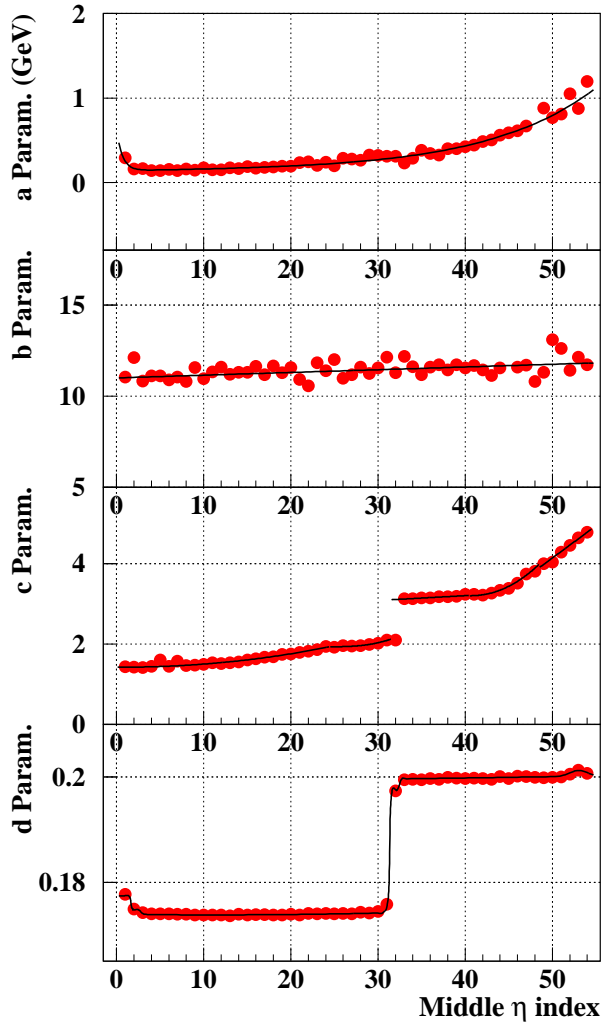}
\caption{\it Calibration parameters used to reconstruct the cluster
energy. In the first and second upper figures the constant and linear
coefficients respectively of the parametrization of the energy
deposited before and within the presampler are shown. In the second
lowest figure the parameters of the energy reconstruction between the
strips and the presampler are illustrated. The last figure represents
the accordion sampling fraction. The functional parametrization of all
the parameters is also shown.}
\label{fig:CaloParameters}
\end{center}
\end{figure}

The parameters of the shower energy reconstruction are derived for all
positions in $\eta$ in steps with the granularity of one middle
cell. The parameters are derived by fitting the energies measured
within the cluster with respect to the energies within the complete
physical volume of the setup. This allows to take automatically into
account a rather large lateral leakage typically amounting to
$\sim$5~\%\ of the shower energy outside the cluster volume. The
parameters of Eq.~\ref{fullparam} found are displayed as a function of
the middle cell $\eta$ indices in Fig.~\ref{fig:CaloParameters}. These
parameters are obtained as follows.

\begin{itemize}
\item[-] $a_\eta$ and $b_\eta$: are fitted using the distribution of
total energy deposited prior and in the presampler versus the energy
measured in the presampler within the cluster. As expected, the offset
$a_\eta$ scales with the amount of material and the sampling $b_\eta$
parameter is almost constant.
\item[-] $c_\eta$: is fitted using the distribution of total energy
deposited between the presampler and the strips compartment of the
accordion versus the square root of the product of energy measured in
the presampler and the strips (corresponding to the geometrical
average of these two energies).
\item[-] $d_\eta$: is fitted using the distribution of the total
energy deposited in the accordion versus the measured energy in all
compartments of the accordion within the electromagnetic EM cluster.
\end{itemize}

The discontinuity in the parameters $c_{\eta}$ and $d_{\eta}$ is due
to the difference in lead thickness between the electrodes A and
B. Apart from the discontinuity at the transition, the accordion total
sampling fraction is mostly constant.


The variation of the $c_{\eta}$ parameter reflects the non trivial
interplay of two effects: the design of the strips was made with a
constant longitudinal extent in units of radiation lengths
corresponding to about 4$X_0$ and the design of the presampler which
has a constant thickness and thus has an increasing depth in units of
radiation lengths.

\subsubsection{Longitudinal Leakage Energy Parametrization}
\label{sec:EneLeakage}

\begin{figure}[htbp]
\begin{center}
\includegraphics[width=10cm]{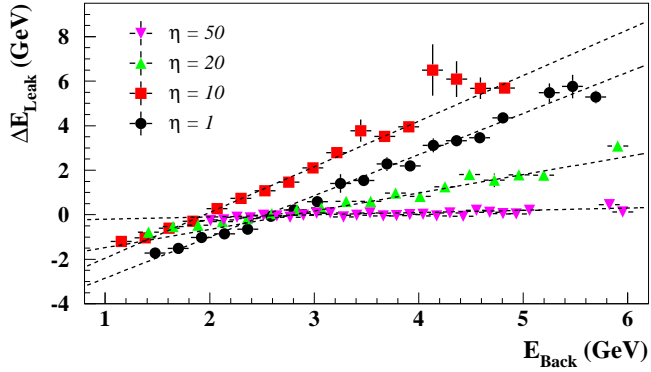}
\caption{\it Difference between actual and average leakage energy as a
function of the energy deposition in the back compartment for various
electron impact points.}
\label{fig:LeakEB}
\end{center}
\end{figure}

\begin{figure}[htbp]
\begin{center}
\vspace{-3.5cm}
\includegraphics[width=9cm]{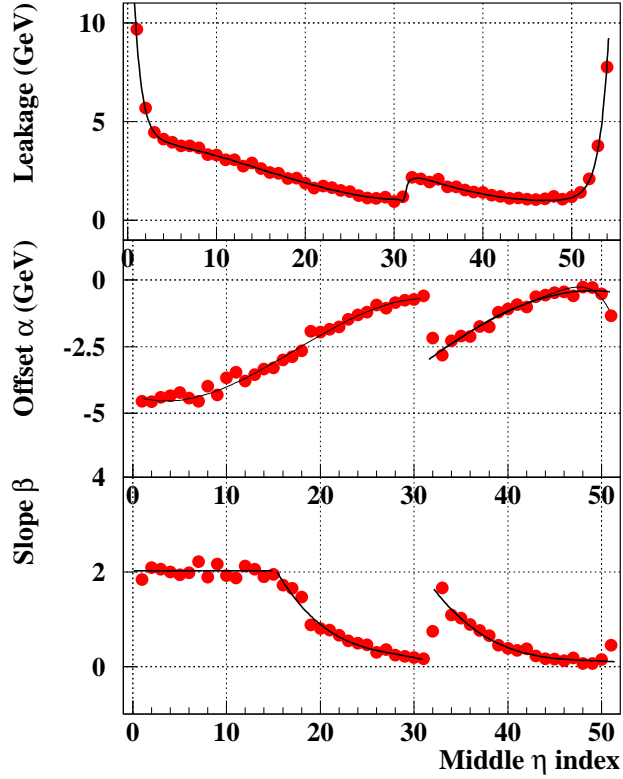}
\caption{\it Calibration parameters used to reconstruct the leakage
energy. The upper plot illustrates the average leakage energy as a
function of pseudo rapidity. The second and third plots show the
variations in $\eta$ of the constant ($\alpha$) and linear ($\beta$)
coefficients used to reconstruct the leakage energy on an
event-by-event basis using the measured energy in the back
compartment.}
\label{fig:LeakageParameters}
\end{center}
\end{figure}

Depending on the $\eta$ coordinate of electrons impinging at 245~GeV
on the modules the amount of leakage can reach non negligible
values. However, these amounts are rather small in regions far from
the edges of the modules. A simple correction corresponding to adding
the average value of the expected leakage energy as a function of the
location of point of impact would be sufficient in regions where the
average leakage does not exceed a few GeV. However, in order to
optimize the energy resolution in the regions near the module edges,
the correlation between the energy deposited in the back compartment
of the accordion and the energy lost longitudinally can be
exploited. The correlation between the longitudinal leakage energy and
the energy deposited in the Back compartment of the accordion is shown
in Fig.~\ref{fig:LeakEB} where the leakage energy is represented as
the difference between the leakage and its average value for a given
pseudo rapidity. The average leakage is shown in the upper plot of
Fig.~\ref{fig:LeakageParameters}.

The average values of the longitudinal leakage are parametrized as a
function of the pseudo rapidity of the impact position of the electron,
as measured by the strips. The expected distributions of the energy
leakage as a function of the energy measured in the back compartment
are fitted for all pseudo rapidities with a granularity of one
middle accordion cell as follows:

$$\Delta E_{Leak}(\eta) = E_{Leak}(\eta)-<E_{Leak}>(\eta) = \alpha_\eta +
\beta_\eta E_{Back}(\eta)$$

The $\beta_{\eta}$ factor represents the approximately constant ratio
of energy leaked longitudinally and the energy deposited in the back
compartment. The $\alpha_{\eta}$ term can be interpreted as the
weighted average energy in the back compartment {\em i.e.}
$\alpha_{\eta} = - \beta_{\eta} <E_{Back}>$.

The results of these fits are also parametrized as functions of the
pseudo rapidity. The leakage is assessed analytically as follows:
$$\xi(E_{Back}^{meas})=<E_{Leak}>(\eta)+\alpha_\eta +
\beta_\eta E_{Back}(\eta)$$
It is then simply added to the shower energy to form the raw
reconstructed energy.

In this procedure the leakage energy depends largely on the energy
deposited in the back compartment of the calorimeter. Since the
leakage correction is evaluated from the Monte Carlo, it is crucial
that the cross talk between middle and back as described in
Section~\ref{sec:crosstalks} be well understood. Because all measured
values of this effect are compatible, it is corrected for
in the Monte Carlo. The leakage correction can then be directly
applied to the data.

The linear correlation between energy deposition in the back
compartment and the leakage energy is manifest. However, at low values
of energy deposition in the back compartment the leakage energy is
systematically above that expected from a pure linear
correlation. This effect arises from events where an early hard
hadronic photon-nucleus interaction occurs within the showering
process. The produced hadronic particles escape the volume of the
electromagnetic calorimeter implying a large longitudinal leakage
without depositing any significant amount of energy in the back
compartment. Such events are relatively rare but when they occur they
carry a lot of energy outside the calorimeter. In this analysis those
events presenting a very large leakage due to photon-nucleus
interaction are removed by the pion hadronic veto. The events with a
small though non negligible fraction of leakage from a photon-nucleus
interaction could not be properly treated in our test beam
protocol. Nevertheless a correction could be designed for ATLAS where
the hadron calorimeter could catch the hadronic tails of electrons
undergoing hadronic interactions.

\subsection{\bf Energy Reconstruction Scheme Performance}
\label{sec:EneRecoSchemePerf}

The application of the described energy reconstruction scheme to the
Monte Carlo simulation for electrons impinging a single cell of the
middle compartment results in a non-uniformity of 0.10~\%. The
uncertainty due to limited statistics of the Monte Carlo samples
implies a non uniformity of about 0.05~\%. The expected systematic non
uniformity arising from the Monte Carlo parametrization thus amounts
to 0.09~\%.

\subsection{\bf Comparison of Data to Monte Carlo}
\label{sec:dataMCcomp}

As mentioned in Sec.~\ref{sec:muA2GeV}, the conversion factor
$f_{I/E}$ from current to energy is estimated from a comparison of 
data and the Monte Carlo simulation. The presampler and the accordion
are normalized independently for each electrodes A and B. For the 2001
running period the normalizations are evaluated from a comparison
between M10 data and the Monte Carlo simulation. For the 2002 running
period the normalization is estimated using the P13 data only. The
constants derived from the comparison with the module P13 are also
applied to the P15 module. When comparing the ratio of the constants
derived from the data between electrodes A and B to those directly
inferred from first principles, as described in
Sec.~\ref{sec:muA2GeV}, in the simulation the difference amounts to
less than 1~\%.

\begin{figure}[htbp]
\begin{center}
\includegraphics[width=10cm]{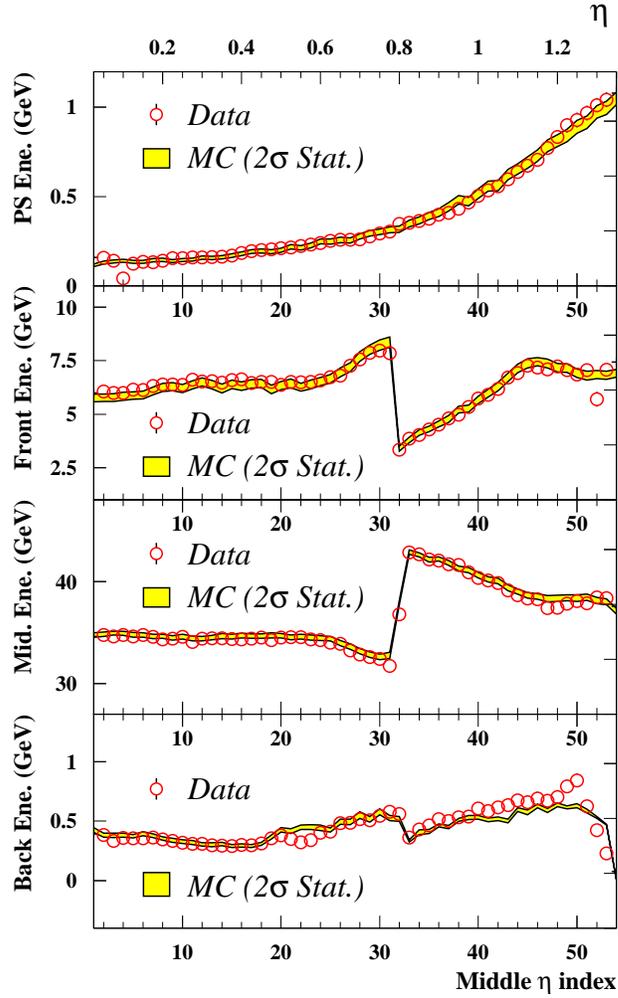}
\caption{\it P15 data versus Monte Carlo comparison in the $\eta$
direction for each compartment individually: the presampler (PS);
strips (front), Middle (Mid.) and Back. The comparison is made at the
constant $\phi$ middle cell index of 11.}
\label{fig:dataMCcomp}
\end{center}
\end{figure}

To illustrate the performance of the simulation
Fig.~\ref{fig:dataMCcomp} displays a comparison of the visible
energy in the data of the module P15 with the Monte Carlo for each
individual compartments of the calorimeter as a function of the $\eta$
middle cell index. A general very good agreement is observed. In
particular, it can be noted that:
\begin{itemize}
\item[(i)] The overall energy calibration of the module P15 is in good
agreement with that of the module P13 and moreover the
inter-calibration of the electrodes A and B is also well reproduced.
\item[(ii)] The good agreement in the absolute scale of the strips
results from a sound treatment of the large capacitive cross talk.
\item[(iii)] The overall acceptable agreement in the back compartment
illustrates that cross talk between the middle and back compartments
has been correctly taken into account.
\end{itemize}

\subsection{\bf Cluster Level Corrections}

To improve the accuracy of the energy reconstruction there are three
further effects that should be taken into account. The first one is
the cluster energy dependence on the EM particle impact position
within one middle cell. Along the $\eta$ and $\phi$ directions
although the transverse leakage is corrected for there is a residual
energy modulation effect due to the limited extent of the
cluster. Along the $\phi$ direction there is an additional energy
modulation due to the structure of the interleaved accordion
absorbers. These two effects are taken into account in the correction
factor $f_{CI}(\eta,\phi)$. The second effect is the energy loss in
the electrodes transition region. It is corrected using the factor
$f_{TR}(\eta)$.  The third effect is the systematic variation in the
electronic calibration due to differences in calibration cable
lengths. It is corrected by means of the factor $f_{Cables}(\eta)$.
The final EM particle energy reconstruction scheme can be written as
follows:

\begin{eqnarray*} E^{rec} = (E^{rec}_{raw} \times f_{CI}(\eta,\phi) \times f_{TR}(\eta) \times f_{Cables}(\eta))
\end{eqnarray*}

\subsubsection{Energy Modulation Corrections}
\label{sec:clustercor}

\begin{figure}[htbp]
\begin{center}
\includegraphics[width=12cm]{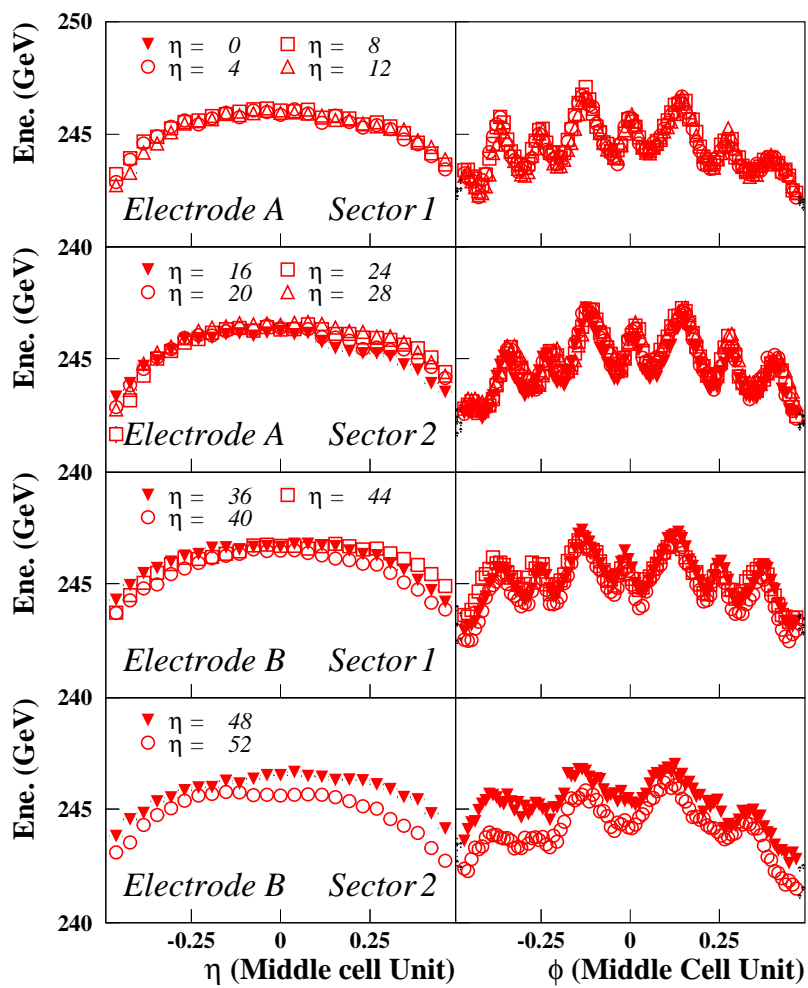}
\caption{\it Energy modulations as a function of the $\eta$ (left
plots) and $\phi$ (right plots) directions for different regions
covering middle cell indices in the $\eta$ direction and integrating
four cells units both in the azimuthal and pseudo rapidity
directions.}
\label{fig:modulations}
\end{center}
\end{figure}

In order to correct for the energy modulations using data, a precise
and unbiased measurement of the impinging electron impact position is
necessary. The $\eta$ position is accurately given by the energy
deposition in the strips. A simple weighted average yields an accurate
estimation of the pseudo rapidity of the electron impact on the
calorimeter. In the azimuthal direction the finest granularity is
given by middle cells and is eight times coarser than that of strips
along the $\eta$ direction. Given the exponentially slender electron
shower profile the azimuthal coordinate given by the weighted average
of the energy depositions in the middle compartment is biased towards
the center of the cell. This bias is referred to as S-shape alluding
to the shape of the distribution of the reconstructed position with
respect to the original one. It is also present along the $\eta$
direction but can be neglected in these studies. Advantage is
therefore taken of the precise position measurement of the wire
chambers located in the test beam to evaluate this S-shape in order to
re-establish an unbiased estimate of the azimuthal coordinate of the
cluster. An energy distribution as a function of $\eta$ and $\phi$ is
evaluated for each middle cell of the P13 module. These distributions
correspond to sliding averages of five cells in $\eta$ in order to
accumulate enough statistics to precisely fit the shape of the
modulations. All cells are fitted and the evolution of the fit
coefficients are parametrized. The modulation correction is therefore
fully analytical.  All spectra in $\eta$ and $\phi$ are displayed in
Fig.~\ref{fig:modulations} and grouped into four regions in $\eta$.

The energy modulations in $\eta$ are parametrized by a parabola as
follows:

\begin{eqnarray}
E_{\eta-corr.}(\eta)=E^{rec}_{raw} / \left[1+C_1(\eta-\eta_C)+C_2(\eta-\eta_C)^2\right]
\end{eqnarray}
where $E^{rec}$ is the raw electron reconstructed energy, $\eta_C$ is
the coordinate of the maximum of the parabola. $C_2$ is the curvature
of the parabola. It is directly linked to the amount of lateral
leakage. $C_1$ is a linear term that introduces an asymmetry in the
energy distribution as a function of $\eta$. This asymmetry is
expected given the cells geometry. The latter term is always very
small. All coefficients are then parametrized using simple functional
forms throughout the module.

When the aforementioned S-shape correction is applied, the energy
modulations become consistent with an expected two-fold modulation
with lengths in middle cell units of $1/4$ and $1/8$ folded with a
parabola resulting from the finite size of the cluster.  As shown in
Fig.~\ref{fig:modulations} all modulations in the barrel are observed
and they display common features. The energy is thus corrected in the
following way:

\begin{eqnarray*}
E_{\phi-corr.}(\phi_{abs}) =
 E^{rec}_{raw}/[(1+C_1(\phi_{abs}-\phi_{C})+C_2(\phi_{abs}-\phi_{C})^2)
 \\ \times a_{8\pi}(\cos{8\pi(\phi_{abs}-\phi_{C})}+ a_{16\pi}(\cos{
 16\pi(\phi_{abs}-\phi_{C})})]
\end{eqnarray*}

$\phi_C$ is the azimuthal coordinate of parabola's maximum. The
coefficients $C_1$ and $C_2$ are the parameters of the parabola. The
linear term is again negligible.  $a_{8\pi}$ and $a_{16\pi}$ are the
amplitudes of the $1/4$ and $1/8$ modulations respectively. All
coefficients are also parametrized using simple functional forms
throughout the module.

As can be seen in Fig.~\ref{fig:modulations} the modulations are less
pronounced in the cells placed at larger pseudo rapidities. This
effect is likely due to the mechanical positioning of the module in
the test beam. A slight deviation from a projective beam
impact on the module can produce such an effect.

When the modulation corrections are applied, an improvement in the
overall energy resolution of typically 30~\%\ is observed.

\subsubsection{Lead transition reconstruction}
\label{sec:leadtransition}

Due to the cylindrical geometry of the barrel calorimeter the sampling
frequency decreases with pseudo rapidity. In order to balance the
energy resolution each module consists of two parts with two separate
electrodes and different lead thicknesses. The lead thickness at high
pseudo rapidity is smaller in order to increase the sampling fraction
and the sampling frequency, given that the geometry is unchanged.
Unfortunately the transition
between the electrode A at low pseudo rapidity and electrode B
at higher pseudo rapidities involves a small uninstrumented region of
roughly 2-3~mm.

\begin{figure}[htbp]
\begin{center}
\includegraphics[width=10cm]{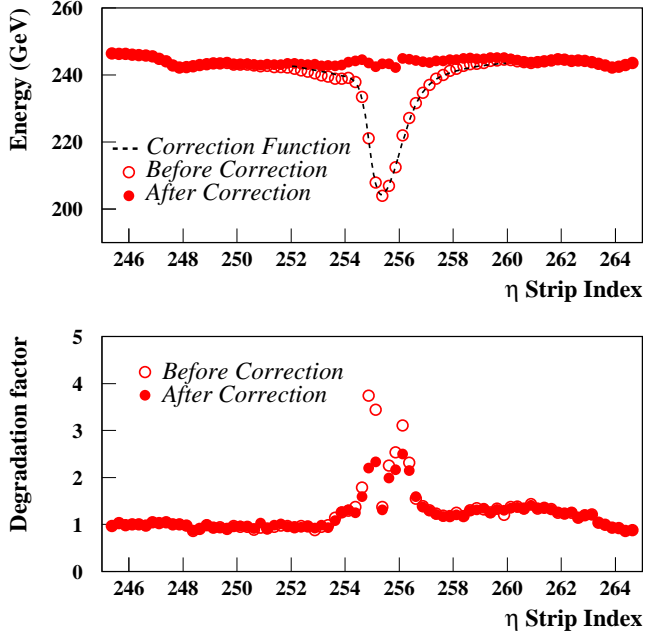}
\caption{\it In the upper figure the variations of the measured energy
across the transition gap between electrodes A and B before and after
the correction is applied are illustrated. The functional form used to
correct for the energy loss in the transition is also shown. The lower
plot illustrates the degradation of the energy resolution with respect
to the neighboring cells throughout the transition.}
\label{fig:barrel_transition}
\end{center}
\end{figure}

In order to study in detail this transition, special high statistics
runs were taken with the P13 module with electrons uniformly covering
the transition region. The average energy evaluated from a Gaussian
fit to the fully corrected energy distribution in $\eta$-bins of a
quarter of one strip cell unit as a function of the pseudo rapidity is
shown in Fig.~\ref{fig:barrel_transition}. Due to the discontinuity in
the lead thickness the sampling fractions are also discontinuous,
therefore in order to have a continuous energy distribution across the
transition, electron sampling fractions need to be applied at the cell
level.

At the transition, the electron energy measurement can be
underestimated by up to $\sim20$~\%. The width of the energy gap is
approximately three strip cells large. A double Fermi-Dirac function
added to a Gaussian is used to fit the energy loss as shown in
Fig.~\ref{fig:barrel_transition}.  In this figure the deterioration
factor ($\sigma_T/\sigma_S$, where $\sigma_T$ and $\sigma_S$ are the
resolutions at the transition and in the neighboring middle cells) in
energy resolution is also shown. After the correction is applied, the
energy resolution is degraded by a factor $\sim2$ over a range of
three strip cells units. However the region where the energy
resolution is deteriorated is very small and corresponds to
approximately 0.003 ($\sim$1 strip cell) in units of pseudo rapidity.

The asymmetry in the loss of energy is due to the increase in bending
angle with depth to maintain a constant gap. The electrodes were
designed to account for this effect and the gap distortion was
unfolded in the pre-bending design. That was not the case for the
absorbers which present a curved transition after bending. In the
Monte Carlo this effect was taken into account and the gap width was
tuned to 2.5~mm on the data. This value corresponds precisely to the
one measured on the modules.

\subsubsection{Cable Length Correction}

The calibration cables carrying the signal from the calibration boards
to the motherboards have different lengths. The resulting variation in
the input signal attenuation is corrected for with the factor
$f_{Cables}(\eta)$ which is evaluated from detailed measurements of
the cable lengths. Because the attenuation occurs before the
calibration signal injection it introduces a small bias in the
calibration procedure. The correction is made {\it a posteriori} at
the cluster level in each layer. It amounts to $\sim$~1.5\%\ in
average and its variations with pseudo rapidity are small (of the
order of one per mil).




\subsection{\bf Modules Uniformity}
\label{sec:modulesunif}

\begin{figure}[hbtp]
\begin{center}
\includegraphics[width=11cm]{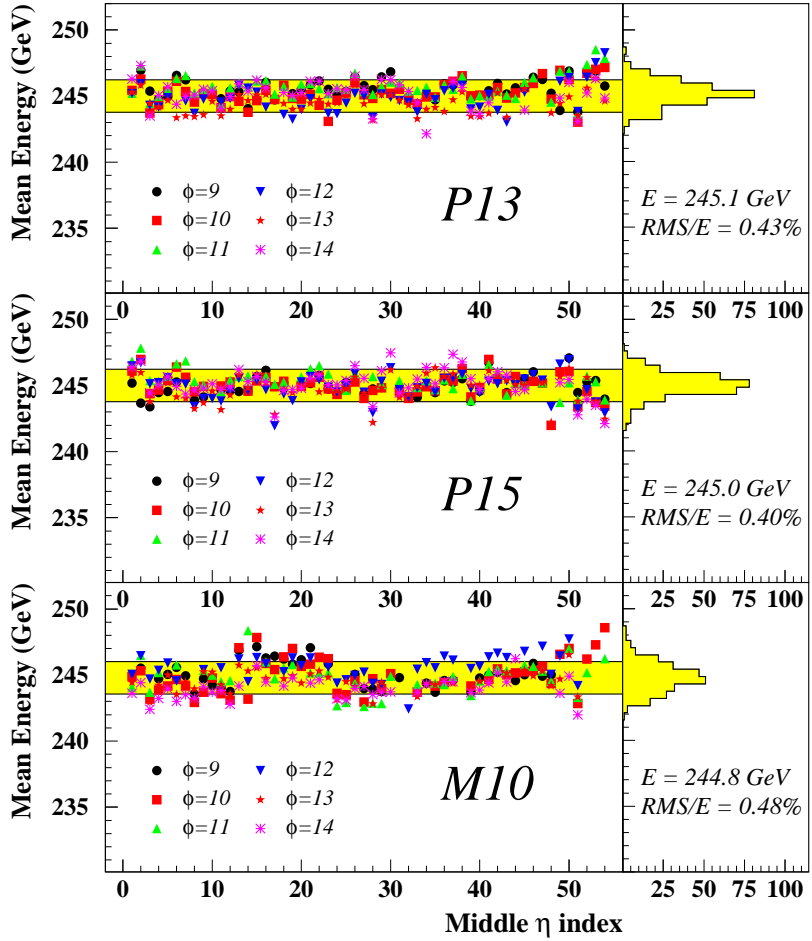} \\
\caption{\it Summary of the energy measurements for each barrel module
for all cells of the FT0 as function of pseudo rapidity. The overall
distribution is shown and the corresponding average and dispersion
values are indicated. The bands corresponds to twice the dispersion of
the measurements.\label{fig:barrel_uniformity}}
\end{center}
\end{figure}

When all corrections are applied the energy distribution for each run
corresponding to one middle cell unit for all barrel modules is fitted
with a Gaussian form starting from -1.5$\sigma$ off the mean value to
determine both the average energy and resolution. All problematic
cells described in Sec.~\ref{sec:clustering} are excluded. 
The mean energies resulting
from the fits to the energy distribution corresponding to the FT0 of
all barrel modules as a function of $\eta$ and for all $\phi$ values
are shown in Fig.~\ref{fig:barrel_uniformity}.


\begin{table}[htbp]
\vspace*{0.5cm}
\begin{center}
\begin{tabular}{|c||c|c|c|}
\hline 
 &  \multicolumn{3}{c|}{$\eta$ Range} \\ \hline
 Module  & Overall [1-54] & Electrode A [1-31] & Electrode B [32-54] \\
\hline 
M10  & 0.48$\pm$ 0.03~\%\ & 0.48$\pm$ 0.03~\%\ &  0.45$\pm$ 0.03~\%\  \\
\hline 
P13  & 0.43$\pm$ 0.03~\%\ & 0.35$\pm$ 0.03~\%\ & 0.48$\pm$ 0.03~\%\  \\
\hline 
P15  & 0.40$\pm$ 0.03~\%\ & 0.36$\pm$ 0.03~\%\ & 0.43$\pm$ 0.03~\%\ \\
\hline 
\end{tabular}

\vspace*{0.5cm}
\caption{\it Non-uniformity expressed in terms of RMS values of the
dispersion of the average energies in the FT0 domain, overall and for
each electrode A and B independently. The statistical uncertainties
are also displayed.}
\label{tab:barrel_uniformity}
\end{center}
\end{table}
\vspace*{0.5cm}

A measure of the non-uniformity at the granularity level of one middle
cell is given by the dispersion (RMS/$<E>$) of the measured
averages. The values obtained are summarized in
Table~\ref{tab:barrel_uniformity}.

Non uniformities of the calorimeter response are typically of the
order of one half percent.

\begin{figure}[htbp]
\begin{center}
\includegraphics[width=11cm]{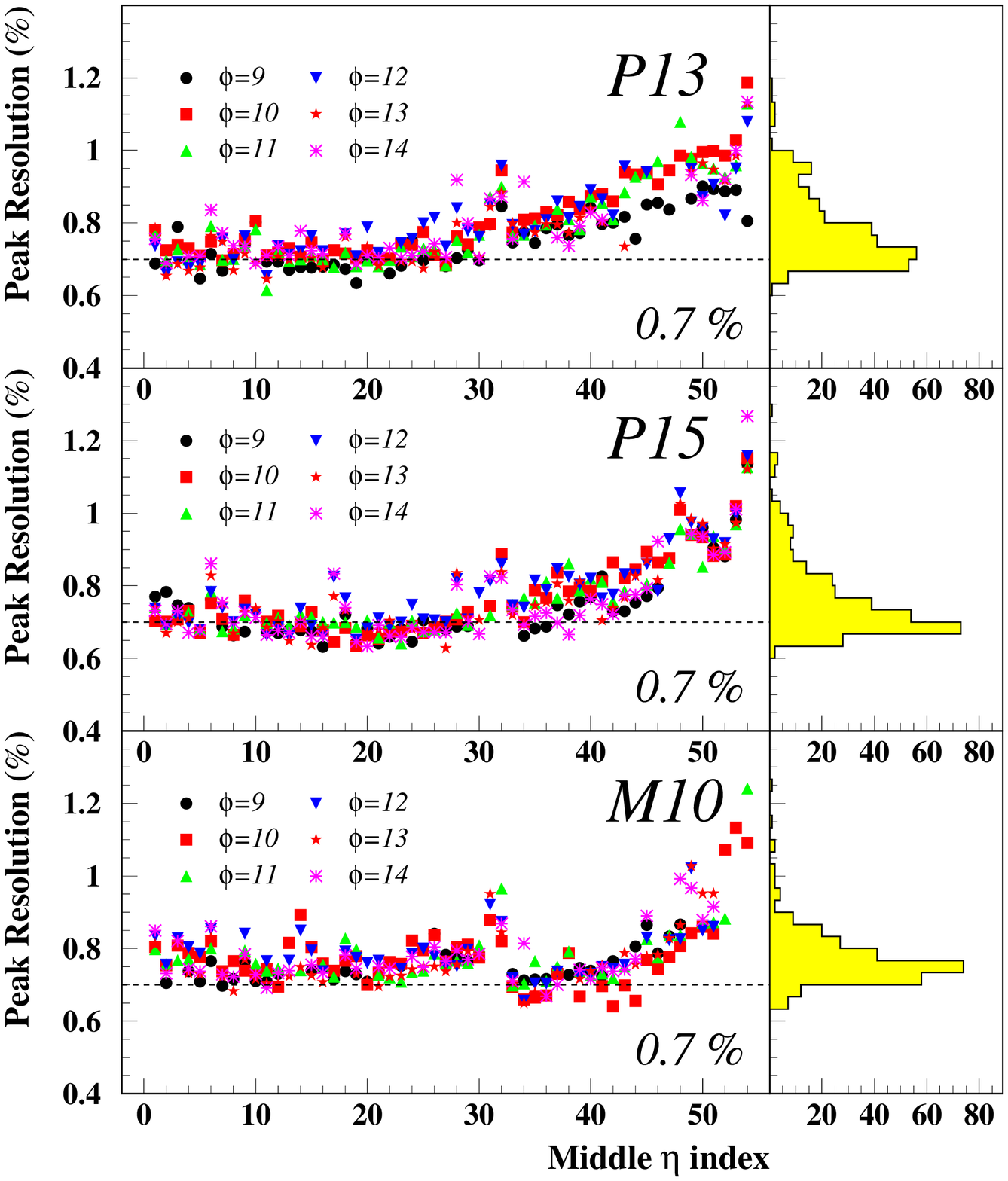}
\caption{\it The local resolution of the cell energy measurements for
all modules and all cells as a function of pseudo rapidity.}
\label{fig:barrel_resolution}
\end{center}
\end{figure}

The widths of the aforementioned Gaussian energy fits as a function of
$\eta$ and for all values of $\phi$ pertaining to the FT0 are shown in
Fig.~\ref{fig:barrel_resolution}. As can be seen in
Fig.~\ref{fig:barrel_resolution} in module M10 around the middle cell
$\eta$ index of 14 and in the modules P13 and P15 around the index 48,
a degradation of the resolution is observed. These effects are due to
bad presampler cells. An overall better energy resolution is obtained
in the module P15, although the modulation corrections were derived
from the module P13. The origin of this difference cannot be easily
traced back. Matter effects could be responsible for differences in
local energy resolution between the module P13 and P15 either at the
level of the constant term or due to differences in the stochastic
term. Differences in manufacturing quality could produce such an
effect.

The general increasing trend of the resolution with respect to pseudo
rapidity is in part due to the increase of the stochastic term and in
part due to the increase of material upstream.

As expected, at the electrode transition where the absorber thickness
changes, a discontinuity in the energy resolution is observed.

\subsection{\bf Relative Energy Scale}

The P13 and P15 modules were exposed to the test beam with the same
calibration and readout electronics and with the same upstream
material. The absolute energy scales for these modules should
therefore be very close or it could imply that a problem at the
construction level occurred. The average reconstructed energy values
for the P13 and P15 modules are 245.1$\pm$0.06~GeV and
245.0$\pm$0.05~GeV, respectively corresponding to a relative
difference of less than 0.1~\%. This global non-uniformity
contribution to the overall energy resolution is small compared to the
non-uniformities observed within each module.

\subsection{\bf Uncorrelated Non Uniformities}

\begin{figure}[htbp]
\begin{center}
\includegraphics[width=12cm]{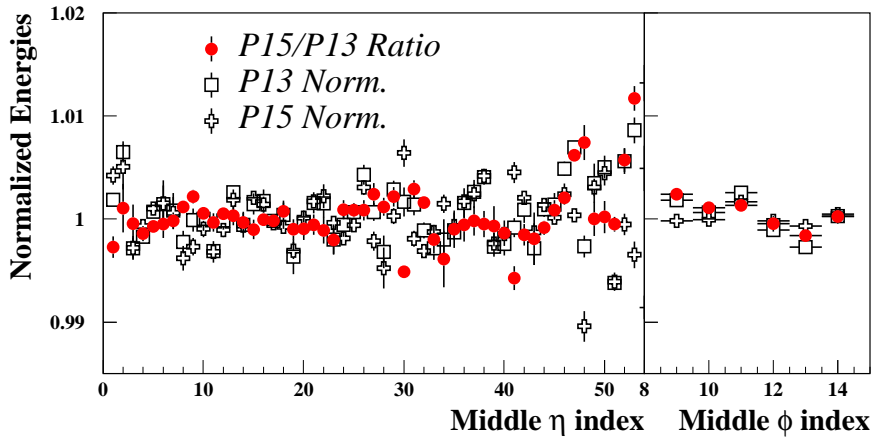}
\caption{\it Profiles of the normalized energy distributions as a
function of pseudo rapidity and azimuthal angle for the modules P13
and P15 and their ratio. These profiles are obtained by averaging over
the FT0 cells in $\phi$.\label{fig:barrel_comparison}}
\end{center}
\end{figure}

In order to disentangle the correlated non uniformities from the
uncorrelated ones, for each cell of the scan the ratio of the average
energies is computed. The distributions plotted in
Fig.~\ref{fig:barrel_comparison} are averages in $\phi$ and in $\eta$
of the mean energy in each cell and their ratio. The uniformity of the
ratio corresponds to the combination of the non correlated
uniformities of the two modules. The dispersion of the distribution of
the ratios amounts to 0.30~\%. The variations in azimuthal angle are
small as shown in Fig.~\ref{fig:barrel_comparison}.

\subsection{\bf Correlated Non Uniformities}

Correlated non uniformities are typically due to the reconstruction
method, inaccuracies in the Monte Carlo simulation or inaccuracies in
the energy corrections. Since the P13 and P15 modules used the same
front-end and calibration electronics their related non-uniformities
are accounted for as correlated.

\begin{table}[thbp]
\vspace*{.3cm}
\begin{center}
\begin{tabular}{|c|c|c|c|c|}
\hline 
Module  & Section & Total & Correlated & Non Correlated  \\
\hline
& Overall & 0.43~\%\ & 0.34~\%\ & 0.26~\%\ \\ 
P13 & Electrode A & 0.35~\%\ & 0.29~\%\ & 0.20~\%\ \\ 
& Electrode B & 0.48~\%\ & 0.34~\%\ & 0.34~\%\ \\ \hline
& Overall & 0.40~\%\ & 0.34~\%\ & 0.21~\%\ \\ 
P15 & Electrode A & 0.36~\%\ & 0.29~\%\ & 0.21~\%\ \\ 
& Electrode B & 0.43~\%\ & 0.34~\%\ & 0.26~\%\ \\ \hline 
\end{tabular}

\vspace*{.3cm}
\caption{\it Details of the correlated and non correlated
non-uniformities for the P13 and P15 modules. The values corresponding
to the electrode A and B are also given separately.\label{tab:NU}}
\end{center}
\end{table}
\vspace*{0.5cm}

Using the dispersion of the ratio of the measured energies for the
modules P13 and P15, the correlated and the uncorrelated contributions
to the uniformity can be separated. The values of correlated and non
correlated non uniformities are displayed in Table~\ref{tab:NU}.

\subsection{\bf Contribution to the Non Uniformity}

All contributions to the non uniformities of the calorimeter response
cannot be easily disentangled. Merging the beam test results with the
quality control measurements, the electronics performance evaluation
and the monte carlo simulation, a non exhaustive list of sources of
non uniformity, displaying the most prominent contributions, is
proposed hereafter.

\subsubsection{Electronics Calibration System}

The calibration system was built within very strict requirements
regarding the precision of the electronics components. The entire
system was thoroughly reviewed and tested. The precision of the three
possible sources of non uniformity (pulsers, injection resistors and
cables) was estimated. All calibration boards were measured on a
test bench and were found to fulfill the requirements. Non
uniformities arising from the calibration system are detailed
in~\ref{sec:CalibrationSystem} and displayed in
Table~\ref{tab:summary}. The overall non uniformities amounts to
$\sim$0.23~\%. This estimate is mostly correlated between the modules
P13 and P15.

\subsubsection{Readout Electronics}

Assuming that the calibration system is uniform, the properly
calibrated readout electronics should not contribute to
non-uniformities. However, small differences in the readout response
could infer second order effects that could imply variations in the
calorimeter response. In order to assess these variations, data was
taken with the P15 module where two middle cells Front End Boards (FEB)
were permuted. The new FEB configuration was calibrated and new data
were taken. The variation in the energy measurement with the board
swap amounted to approximatively 0.1~\%.

\subsubsection{Signal Reconstruction}
\label{sec:enerecoscheme}
For various runs taken on the P15 module the data were reconstructed
using the two methods described in Sec.~\ref{sec:signalreco}. In order
to assess the possible non uniformities arising from the signal
reconstruction method the dispersion of the differences in the average
energy between the two methods is taken. The observed RMS amounts to
0.25~\%. This estimate is likely to be an overestimate of the
intrinsic bias of the method.

\subsubsection{Monte Carlo Simulation}

The full Monte Carlo simulation of the experimental setup cannot
perfectly reproduce the actual data. Non uniformities in the
calorimeter response can thus arise from the simulation. The precision
of the Monte Carlo description will directly impact the energy
measurement. The difference between data and Monte Carlo presented in
Sec.~\ref{sec:dataMCcomp} amounts to 0.08~\%. This figure, derived
from the dispersion of the distribution of the difference in measured
energies between data and Monte Carlo, represents the expected non
uniformity arising from the simulation inaccuracies.

\subsubsection{Energy reconstruction Scheme}
The energy reconstruction scheme involves a large number of
parameterizations and fits. Inaccuracies of these parameterizations
will impact the energy measurements and can induce a non uniform
response. A measure of the inaccuracies of the parametrization is the
residual systematic non uniformity in the Monte Carlo simulation. As
was shown in Sec.~\ref{sec:EneRecoSchemePerf}, this effect amounts to
0.09~\%.

\subsubsection{Module Construction}

The non uniformities related to the construction of the modules are
the dominant source of non-correlated non uniformities. The main sources
of the non-uniformity in the construction of modules are the lead
thickness and the gap dispersion.

\begin{itemize}

\item[(i)] The impact of the variations in lead thickness on the EM
energy measurements was assessed and a scaling factor of 0.6 was found
between the dispersion of the lead thickness and the dispersion of the
EM energies.

\item[(ii)] Similarly the impact of the variations of the gap were
studied and a scaling factor of 0.4 was found between the dispersion
of the gaps and that of the EM energy measurements.

\end{itemize}

From the measurements presented in Sec.~\ref{sec:leadthickness} the expected non
uniformity obtained are displayed in Table~\ref{tab:summary}.

\subsubsection{Modulation Corrections}

The energy modulation corrections can impact the calorimeter response
to electrons at different levels either by affecting the uniformity or
the local constant term.

The modulation corrections were evaluated on the module P13 only and
were then applied to all other modules. For this reason it is
difficult to disentangle the correlated from the non correlated part
of the correction. For the sake of simplicity this effect will be
considered as exclusively non correlated. To evaluate its impact both
on the uniformity and on the local constant term, the complete
analysis is done restricting the measurement to a small region
accounting for 20~\%\ of the cell around its center. The differences
found are of 0.14~\%\ and 0.10~\%\ for the modules P13 and P15
respectively.
 
\subsubsection{Time Stability}

\begin{figure}[htbp]
\begin{center}
\includegraphics[width=10cm]{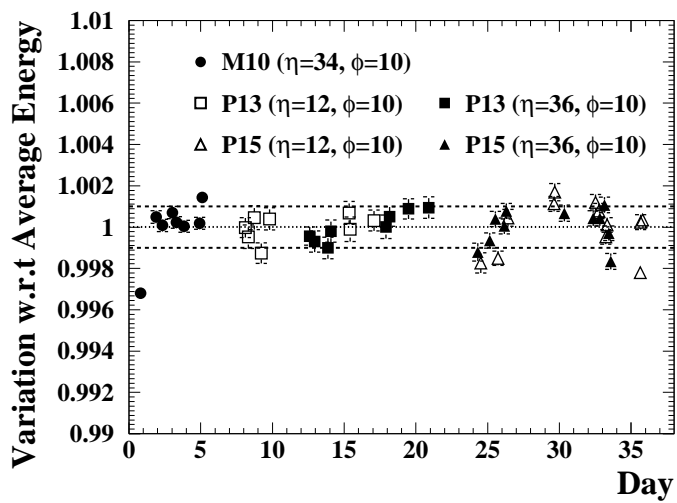}
\caption{\it Energy measurements for two reference cells in modules
P13 and P15 and in module M10, as a function of time. The $\pm
1$\promille\ variation band is also indicated.}
\label{fig:timestability}
\end{center}
\end{figure}

In order to check the stability of the energy reconstruction, reference
cells were periodically scanned with the 245~GeV electron beam. Two
cells were chosen for the modules P13 and P15 both at a middle cell
$\phi$ index of 10 and at $\eta$ indices of 12 and 36. For the module
M10 only one reference cell was taken at an $\eta$ index of 34. The
variation of the energy reconstruction with time is illustrated in
Fig.~\ref{fig:timestability}.  

From the observed variations, the impact on the energy measurements
are estimated to be 0.09~\%, 0.15~\%\ and 0.16~\%\ for the modules P13,
P15 and M10 respectively.

\subsubsection{Summary}

All known contributions to the non uniformity are summarized in
Table~\ref{tab:summary}. The good agreement achieved between the data
and the expectation illustrates that the most sizable contributions to
the non uniformities have been identified. 

\begin{table}[h]
\vspace*{.3cm}
\begin{center}
\begin{tabular}{|c|c|c|}
\hline \hline
Correlated Contributions & \multicolumn{2}{c|}{Impact on Uniformity} \\ \hline \hline
Calibration & \multicolumn{2}{c|}{0.23~\%} \\ \hline
Readout Electronics & \multicolumn{2}{c|}{0.10~\%} \\ \hline 
Signal Reconstruction & \multicolumn{2}{c|}{0.25~\%} \\ \hline 
Monte Carlo & \multicolumn{2}{c|}{0.08~\%} \\ \hline
Energy Scheme & \multicolumn{2}{c|}{0.09~\%} \\ \hline
\hline 
Overall ({\bf data}) & \multicolumn{2}{c|}{0.38~\% ({\bf 0.34~\%})} \\ \hline \hline
Uncorrelated Contribution & P13 & P15 \\ \hline \hline
Lead Thickness & 0.09~\%\ & 0.14~\%\ \\ \hline
Gap dispersion & 0.18~\%\ & 0.12~\%\ \\ \hline
Energy Modulation & 0.14~\%\ & 0.10~\%\ \\ \hline
Time Stability & 0.09~\%\ & 0.15~\%\ \\ \hline
Overall ({\bf data}) & 0.26~\%\ ({\bf 0.26~\%}) & 0.25~\%\ ({\bf 0.23~\%}) \\ \hline \hline
\end{tabular}

\vspace*{0.5cm}
\caption{\it Detail of the expected contributions to the uniformity and
to the constant term. \label{tab:summary}}
\end{center}
\end{table}
\vspace*{0.5cm}

The module P15 displays a slightly better uniformity than the other
modules. None of the control measurements support this
observation. However, as shown in Sec.~\ref{sec:leadthickness} the
granularity of the control measurements was not particularly
high. Manufacturing differences within such granularity may not be
observable but could impact the uniformity.

\subsection{\bf Local and Overall Constant Term}

\begin{figure}[htbp]
\begin{center}
\hspace*{-.6cm}
\includegraphics[width=16cm]{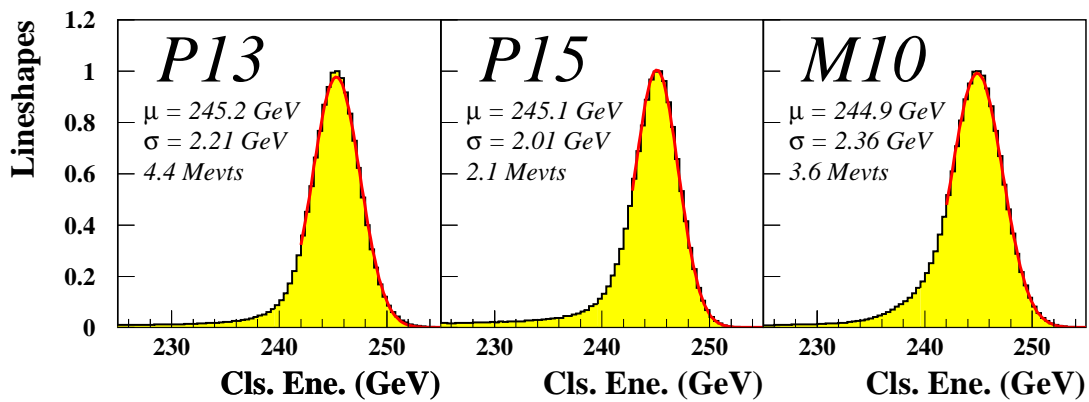}
\caption{\it The energy lineshapes for the barrel modules P13, P15 and
M10 containing respectively 4.4, 2.1 and 3.6 million events. The
simple gaussian fits to energy peak are displayed and the fit
parameters are indicated.}
\label{fig:barrel_lineshapes}
\end{center}
\end{figure}

The distributions of all energy measurements for each barrel module
are shown in Fig.~\ref{fig:barrel_lineshapes}. These distributions
showing the overall energy measurement resolution throughout each
module are fitted with a simple Gaussian form starting from
-1.5$\sigma$ off the mean values.

The corresponding overall resolutions are 0.93~\%, 0.85~\%\ and 0.96~\%\
for the modules P13, P15 and M10 respectively. The main components of
these resolutions are: (i) the non-uniformity of the modules, (ii) the
local constant terms, (iii) the stochastic terms and (iv) the
electronic noise.

The stochastic term (iii) has been precisely measured
in~\cite{linearity} at one fixed point and more broadly estimated over
the full range in pseudo rapidity in~\cite{NIMBARREL0}. Taking into
account the $\eta$-dependence of the stochastic\footnote{The
$\phi$-dependence of the stochastic term is assumed to be negligible
here.} term as derived from~\cite{NIMBARREL0}, the noise as evaluated
from random trigger events, the local resolutions and the beam energy
spread amounting to 0.08~\%\ a local constant term can be derived for
each cell. The distribution of local constant terms yields an average
of 0.30~\%, 0.25~\%\ and 0.36~\%\ for the P13, P15 and M10 modules
respectively. The dispersion of the local constant term is typically
of 0.11~\%\ absolute. The overall constant term for all modules can be
derived from the average local constant terms by simply adding the
measured non-uniformities. The global constant terms obtained are thus
0.52~\%, 0.48~\%\ and 0.60~\%\ for the modules P13, P15 and M10,
respectively. These results are derived under the assumption that the
stochastic term is the same in all modules. A slight variation in the
stochastic term could also explain the differencies observed between
the P13 and P15 resolutions.

The constant terms (average local or global) are considerably better
for the module P15 with respect to the other modules. This observation
supports the hypothesis that the module P15 was better manufactured,
but could also be due a better stochastic term.  All results are
summarized in Table~\ref{tab:unifbarrel}.

\begin{table}[htbp]
\vspace*{.3cm}
\begin{center}
\begin{tabular}{|c||c|c|c|}
\hline 
  & \multicolumn{3}{c|}{Barrel modules}  \\
\hline 
 Module & P13 & P15 & M10\\
\hline 
$<E>$ & 245.1 GeV & 245.0 GeV & 244.8 GeV \\
\hline 
{\bf RMS/}$\mathbf{<E>}$ & {\bf 0.43~\%} &  {\bf 0.40~\%} & {\bf 0.48~\%}  \\
\hline 
$\sigma/\mu$ & 0.82~\%\ & 0.79~\%\ & 0.78~\%\ \\
\hline
{\bf Local Constant Term} & {\bf 0.30~\%} &  {\bf 0.25~\%} & {\bf 0.36~\%}  \\
\hline
{\bf Global Constant Term} & {\bf 0.52~\%} &  {\bf 0.47~\%} & {\bf 0.60~\%}  \\
\hline 
\end{tabular}

\vspace*{0.5cm}
\caption{\it Mean energy, non-uniformity, average energy resolution
and global constant term for the three tested barrel modules over the
entire analysis region.}
\end{center}
\label{tab:unifbarrel}
\end{table}
\vspace*{0.5cm}

\section{ENDCAP UNIFORMITY} 
\label{Sec:Endcap}

\subsection{\bf Energy Reconstruction}
\label{sec:ECreco}

\subsubsection{Clustering Scheme}
\label{sec:ECclustering}

The electron energy is reconstructed as in the barrel by summing the
calibrated energies deposited in the three calorimeter compartments
around the cell with the largest energy deposit in the middle
compartment. The choices of cluster sizes result from the best
compromise between noise and energy leakage. Table~\ref{clustertab}
summarizes the cluster size in the three compartments. It can be noted
that the cluster size changes dramatically in the front compartment
due to the variations in the cell granularity. As was the case in the
barrel electron clusters the choice of one or two strip cells in
$\phi$ relies upon whether the shower develops near the cell center or
not.

\begin{table}[htbp]
\vspace*{0.5cm}
\begin{center}
\begin{tabular}{|l||c|c|c|c||c|}
\hline
    &  \multicolumn{4}{c||}{Outer wheel} & Inner wheel\\
\hline $\eta$-range & [1.5, 1.8] & [1.8, 2.0] & [2.0, 2.4] & [2.4,
 2.5] & [2.5, 3.2] \\ \hline Front  & $23\times1(2)$ & $15\times1(2)$ &
 $11\times1(2)$ & $3\times1(2)$ & --- \\ \hline Middle &
 \multicolumn{4}{c||}{$5\times5$} & $3\times3$ \\ \hline Back &
 \multicolumn{4}{c||}{$3\times5$} & $3\times3$ \\ \hline
\end{tabular}

\vspace*{0.5cm}
\caption{\it Cluster size ($\Delta \eta_{cell} \times \Delta \phi_{cell}$) 
per layer around cell with maximum energy deposit.
\label{clustertab}}
\end{center}
\end{table}
\vspace*{0.5cm}



\subsubsection{Reconstruction Scheme}

As mentioned in Sec.~\ref{Sec:TBsetup}, the barrel and endcap modules
were tested in different beam lines at two different maximum energies
namely 245~GeV and 119~GeV respectively. At this lower beam electron
energy the relative impact of the longitudinal leakage is extremely
small but the effect of inactive material is larger. Another major
difference intrinsic to the detector is the absence of a
presampler. However, the material upstream of the calorimeter is
approximately constant. For these reasons the energy reconstruction
scheme can be considerably simplified at first order in the endcaps
with respect to that used in the barrel. A single normalization factor
for all modules can thus be used to derive the total energy from the
visible energy measured in the liquid argon. It is derived after all
corrections are applied.

Nevertheless two main complications arise: the first one from the
continuous decrease of the liquid argon gap with pseudo
rapidity (since the HV is set on a sector basis, the signal
response will vary with $\eta$);  the second from the effective
variations of the cluster size with the pseudo rapidity. The fraction
of the total 119~GeV electron energy contained in the cluster exceeds
92~\% at high $\eta$ and even more at low $\eta$. A single {\it ad hoc}
correction can be derived from the data to correct for both effects
simultaneously. This correction is described in Sec.~\ref{sec:hvcor}

\subsection{\bf Cell Level Corrections}
\label{sec:ECclustercor}




\subsubsection{{\it Ad hoc} Residual High Voltage and Leakage Correction}
\label{sec:hvcor}

\begin{figure}[htbp]
\begin{center}
\epsfig{file=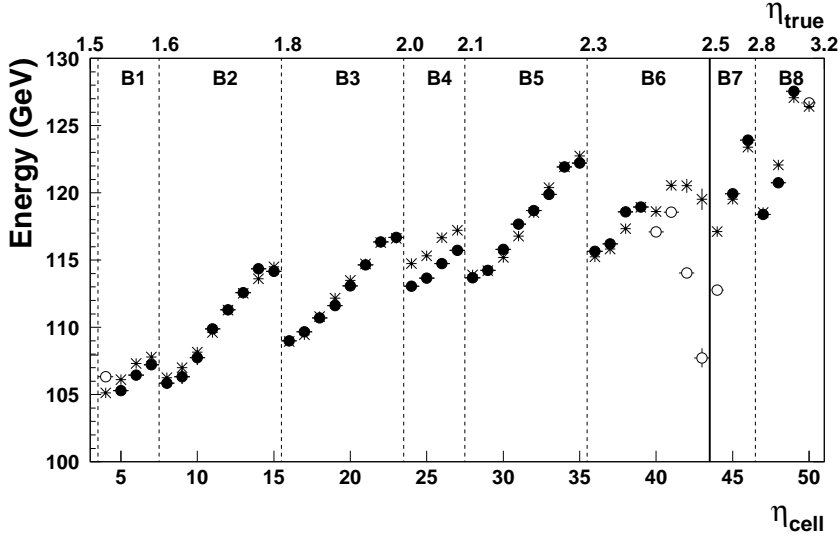,
width=12cm,bbllx=0pt,bblly=0pt,bburx=700pt,bbury=420pt}
\end{center}
\caption{\it Energy averaged over $\phi$ for one module as a function
of $\eta$ before high voltage correction.  The vertical dashed lines
separate the high voltage sectors and the full line at $\eta=2.5$
separates the outer and the inner wheel. The full (empty) dots
correspond to cells inside (outside) the analysis region. The stars
correspond to the full Monte Carlo simulation results.}
\label{HV_mod}
\end{figure}

The fact that the liquid argon gap thickness decreases continuously
along $\eta$ whereas the high voltage changes by steps translates into
a linear increase in signal response with $\eta$, within each high
voltage sector. The transverse leakage will affect the energy in an
opposite manner and {\it a priori} not completely linearly, however
its effect is expected to be smaller.  The overall variation is
illustrated in Fig.~\ref{HV_mod}, where the seven (two) HV sectors of
the outer (inner) wheel are separated with vertical dashed lines.  A
good agreement is achieved with a full Monte Carlo GEANT
simulation\footnote{The 2~\% discrepancy observed in B4 sector is not
understood but could be due to a bad tuning of the HV value of
$\sim50$~V / 1500~V.}~\cite{MC}. The crack between the two wheels
around $\eta=2.5$ distorts the expected linear behavior in sectors B6
and B7.  At this level, the non-uniformity of the response over the
analysis region (RMS of the mean energy distribution) is around 4~\%.

\begin{figure}[htbp]
\begin{center}
\epsfig{file=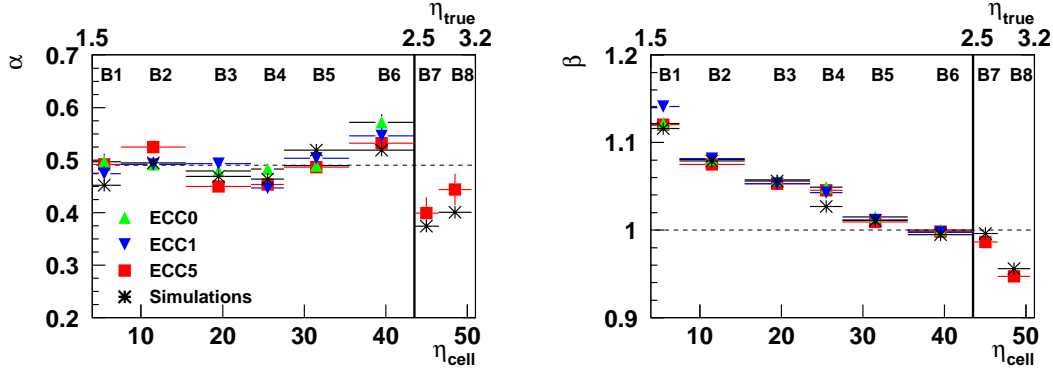,
width=\linewidth,bbllx=10pt,bblly=15pt,bburx=690pt,bbury=270pt}
\end{center}
\caption{\it High voltage correction parameters $\alpha$ (left) and
$\beta$ (right) obtained in the eight scanned HV sectors for the three
tested modules. Results have been averaged over $\phi$.  The vertical
full line at $\eta=2.5$ separates the outer and the inner wheel. The
stars correspond to the full Monte Carlo simulation results.}
\label{HV_mod2}
\end{figure}

These effects are corrected for by weighting the energy of
each cell, depending on its $\eta$-position (taken at its center) and
its HV sector ($l$), by~:
\begin{eqnarray}
E^{cell}_{HV-corr.}(\eta, l)=E^{cell} \cdot \frac{\beta^l}{1+\alpha^l
\cdot (\eta-\eta^l_{center})}
\label{eq:hvcor}
\end{eqnarray}
where $\eta^l_{center}$ is the $\eta$-value at the center of HV sector
$l$.  The coefficients $\alpha^l$ and $\beta^l$ are the correction
parameters of sector $l$, $\beta^l$ being a normalization factor,
close to~1, accounting also for inaccurate high voltage settings. They
are determined by a linear fit. The inner wheel case is more
complicated and the resulting parameters are slightly biased.  The
results obtained for $\alpha^l$ and $\beta^l$ in the eight scanned HV
sectors\footnote{The first HV sector, covering the $\eta$-range
[1.375-1.5], has not been completely scanned (at most 2 cells in
$\eta$) and is not included in the analysis.}  are shown in
Fig.~\ref{HV_mod2} for the three tested modules.

A good agreement between modules is observed and $\alpha$ depends very
weakly on $\eta$. Therefore a single value of $\alpha=0.49$, in good
agreement with the full Monte Carlo simulation, is used for all
modules. 




\subsubsection{Capacitance correction}
\label{sec:capacor}

\begin{figure}[htbp]
\begin{center}
\epsfig{file=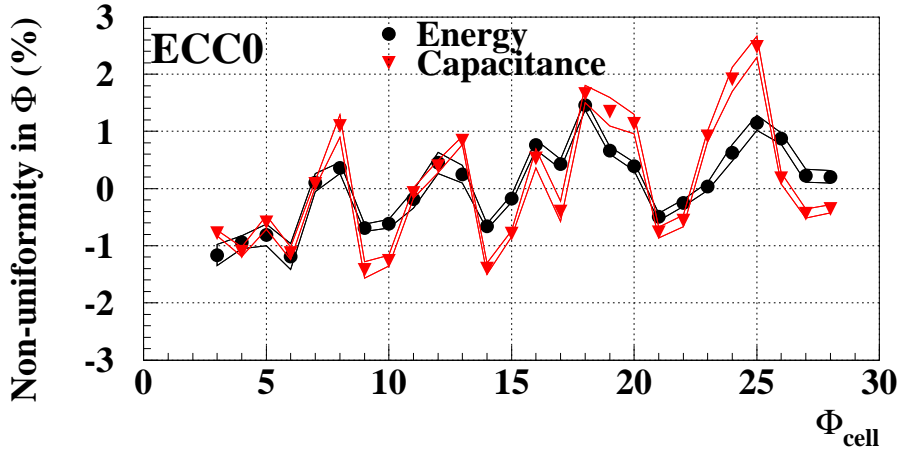,
width=0.82\linewidth,bbllx=12pt,bblly=265pt,bburx=510pt,bbury=520pt}
\end{center}
\caption{\it Variation along the $\phi$ direction of the measured ECC0
outer wheel middle cell capacitance (red triangles) and of the energy
measured in beam test (black points). All points have been averaged
over $\eta$.}
\label{fig:capaecc0}
\end{figure}

Along the $\phi$ direction, the gap thickness is in principle kept
constant by the honeycomb spacers.  However, the energies measured in
the test beam show an unexpected non-uniformity along $\phi$, almost
at the percent level~\cite{notefabrice,rodier}.  This effect can be
correlated with the variations of middle cell capacitance along $\phi$
measured independently, as shown in Fig.~\ref{fig:capaecc0} for ECC0.
The $\phi$-dependence of the energy can thus be explained by local
fluctuations of the gap thickness, generated during the module
stacking.  The effect is almost independent of $\eta$.  Even if it
corresponds to small absolute deviations (the gap thickness is roughly
3~mm at low $\eta$ and 1~mm at high $\eta$, 1~\% represents only a few
tens of microns), it must be corrected in order to achieve the best
response uniformity.

The energy is then corrected by weighing each cell in the following
way~:
\begin{eqnarray}
E^{cell}_{capa-corr.}(\phi)=E^{cell} /
\left(\frac{C_\phi}{<C_\phi>}\right)^\alpha
\end{eqnarray}
where $C_\phi$ is the cell capacitance, $<C_\phi>$ is its average over
all $\phi$ and $\alpha$ the high voltage correction parameter (see
section~\ref{sec:hvcor}).  The effect is assumed to be independent of
the depth. The middle cell measurements are used for all
compartments. The $\alpha$ exponent was empirically chosen but was
found to yield a near to optimal uniformity. It illustrates the
interplay between the high voltage and the capacitance
corrections. Such a correction is not performed in the ECC5 inner
wheel, whose uniformity along $\phi$ is very good.  For ECC1, no
accurate capacitance measurement was made. An {\it ad hoc} correction
is thus extracted from the $\phi$-dependence of the energy averaged
over $\eta$.  As it corrects for local stacking effects, the
correction has to be specific for each cell and each module. In ATLAS
the correction based on the capacitance measurements could be further
refined by inter calibrating $\phi$-slices of the calorimeter with
$Z^0\rightarrow e^+e^-$ events.


\subsection{\bf Cluster Level Corrections}

\subsubsection{Cluster Energy Correction for High Voltage problems}

Another problem, which appeared on ECC1, was that two electrode sides,
which were on spare lines, were not holding the high voltage, inducing
an energy loss depending on the event impact point position in $\phi$
with respect to the faulty electrode.
The dependence of the cluster energy with $\phi$ was parametrized and
corrected with a parabola. The energy resolution of the affected
cells is degraded by $\sim50$~\% and the uniformity in the
corresponding HV zone ($\Delta\eta\times\Delta\phi\sim 0.2\times0.4$)
is degraded by $\sim20$~\%.

\subsubsection{Endcap $\eta$ and $\phi$ energy modulation corrections}

To correct for energy variations along the $\eta$ direction a
parametrization similar to that used in the barrel is
considered. However, as the transverse size of the electromagnetic
shower is constant and the cell dimension decreases with pseudo
rapidity, the leakage is expected to increase with $\eta$, and thus
the absolute value of the quadratic term of the parabola is expected
to increase with pseudo rapidity. This effect is sizable only in EC
modules and a good agreement is achieved among the three tested
modules. The quadratic parameter is linearly parametrized as a
function of $\eta$~\cite{notefabrice}.
 
\begin{figure}[htbp]
\begin{center}
\includegraphics[width=13cm]{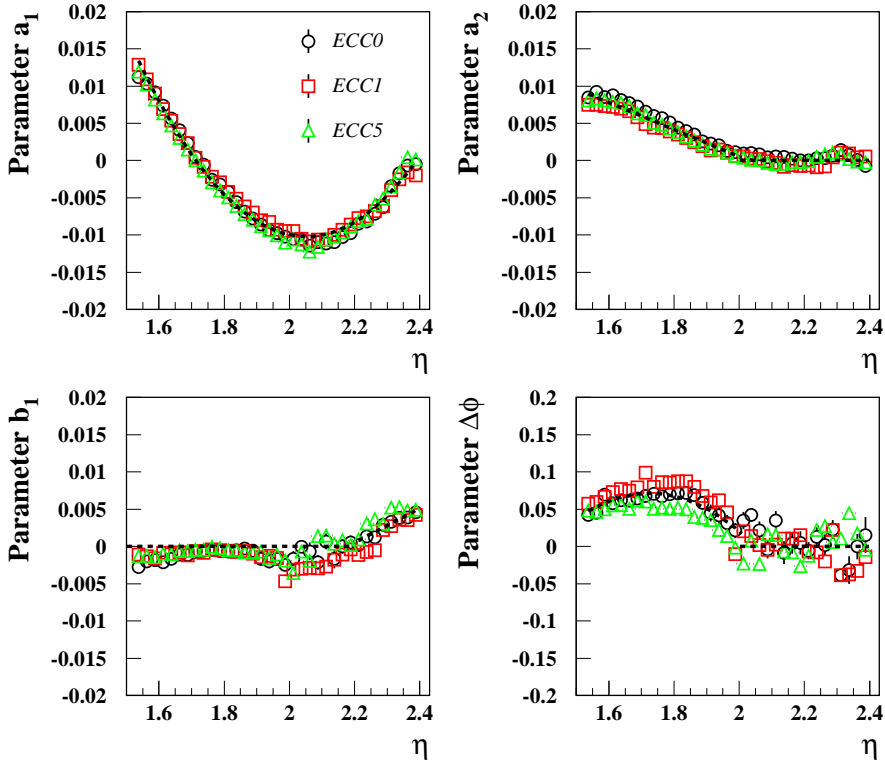}
\end{center}
\caption{\it Coefficients of $\phi$-modulation corrections, averaged
over $\phi$, as a function of $\eta$ for the three tested
modules. Their linear or parabolic parametrisations are superimposed.}
\label{Paramphimod}
\end{figure}

The $\phi$ modulation for EC modules is fitted and corrected for with
the following function~\cite{garcia}~:

\begin{eqnarray}
E_{\phi-corr.}(\phi_{abs})=E / \left(1+\sum_{i=1}^2a_i\cos\left[2\pi i
(\phi_{abs}-\Delta\phi)\right]+b_1\sin\left[2\pi\phi_{abs}\right]\right)
\end{eqnarray}

where $\phi_{abs}$ is in absorber units, $a_1$ and $a_2$ are the
coefficients of the even component (it has been checked that only two
terms are necessary), $b_1$ the coefficient of the odd component one
and $\Delta\phi$ a phase shift. The parameters of the fits are
displayed in Fig.~\ref{Paramphimod} for all three endcap modules. A
good agreement between the fitted parameters for the different modules
is observed. Given the non trivial geometry of the endcap modules this
result highlights the manufacturing quality of the modules. As was the
case for the barrel modules, a single correction derived from a fit of
the coefficients can be used.

\subsection{\bf Modules uniformity}
\label{sec:unif}

\begin{figure}[btph]
\begin{center}
\includegraphics[width=11cm]{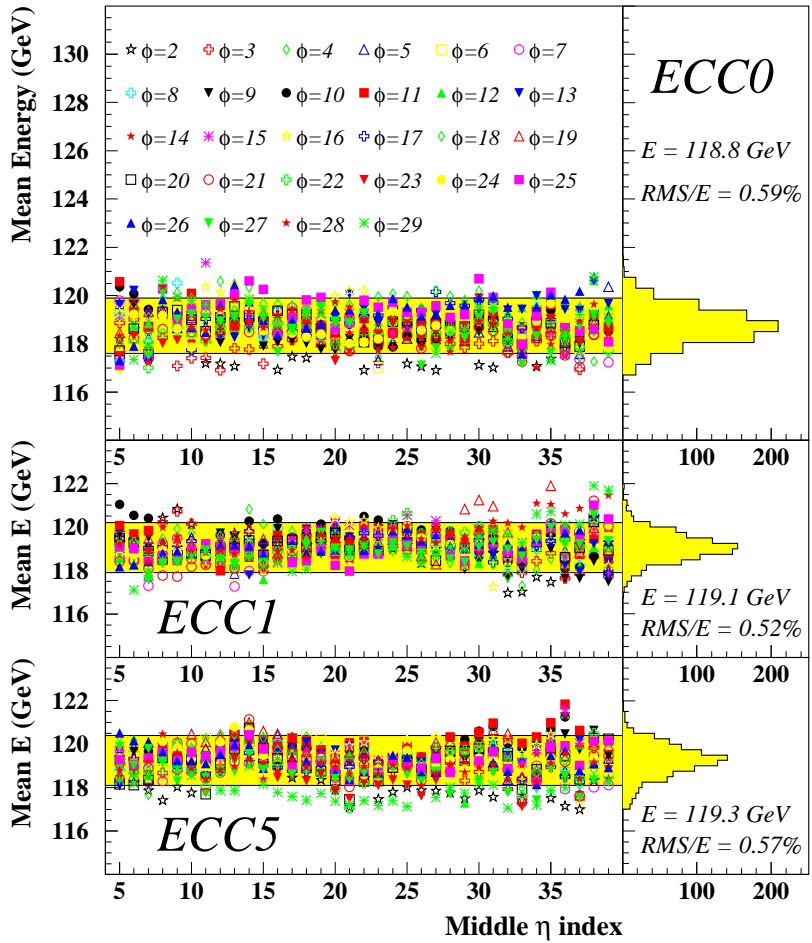}
\end{center}
\caption{\it Mean reconstructed energy as a function of the hit cell
position in $\eta$ for a 119~GeV electron beam. Results are shown for
the outer wheel of all three tested modules.}
\label{fig:unifall}
\end{figure}

The mean energies as reconstructed from a gaussian fit to the energy
distribution after all corrections are shown in Fig.~\ref{fig:unifall}
for all the cells and for the three tested modules. Their dispersion
across the outer wheel analysis region is approximately 0.6~\% for all
modules~\cite{notefabrice}. It is better for ECC1 because an {\it ad
hoc} capacitance correction was used. A similar result is obtained for
the ECC5 inner wheel: the response non-uniformity over the 25 cells
is approximately 0.6~\%.

With the high voltage correction the uniformity is of the order of 1~\%
(0.78~\% for ECC0, 0.86~\% for ECC1 and 0.65~\% for ECC5). The
capacitance correction yields an uniformity close to final
one. Cluster level corrections do not significantly impact the module
response uniformity, but they improve the energy resolutions. The
problematic channels (for instance for high voltage failures in ECC1
and ECC5), that have been kept, do not degrade the results. If they
were excluded, the non-uniformity would improve by less than 0.01~\%.
These results are summarized in Table~\ref{tab:unifEC}.

\subsection{\bf Resolution and Overall Constant Term}
\label{sec:unifEC}

\begin{figure}[htbp]
\begin{center}
\includegraphics[width=11cm]{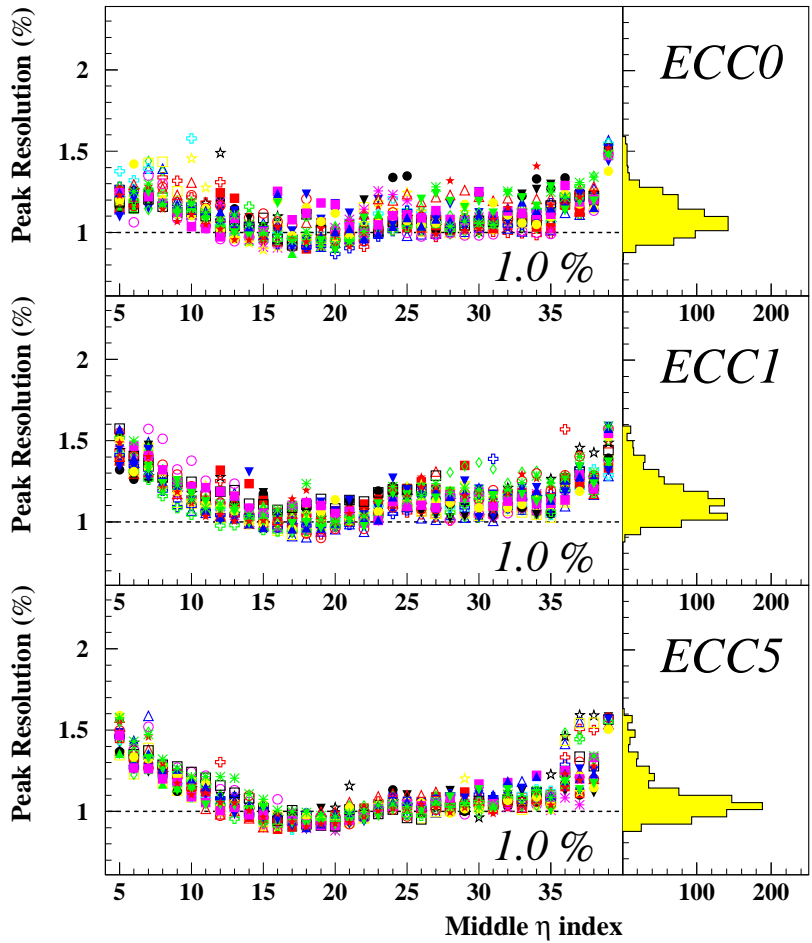}
\end{center}
\caption{\it Reconstructed energy resolutions as a function of the hit
cell position in $\eta$ for a 119~GeV electron beam.  Results are
shown for the three tested modules (outer wheel). The legend used for the different $\phi$ values here is the same as in Fig.~\ref{fig:unifall}}
\label{fig:ResolutionEC}
\end{figure}

The energy resolution as derived from the fit to the energy
distribution for all cells and after all corrections are applied is
shown in Fig.~\ref{fig:ResolutionEC}. When quadratically subtracting
the electronic noise term of $\sim$200~MeV and a beam spread of 0.07~\%\ the
energy resolutions are compatible with those reported
in~\cite{NIMENDCAP0}. Assuming an average local constant term of
0.35~\%\ an average value of the stochastic term of
11.4$\pm$0.3~\%$/\sqrt{E}$ is found.

\begin{table}[htbp]
\vspace{.3cm}
\begin{center}
\begin{tabular}{|c||c|c|c||c|}
\hline 
  & \multicolumn{3}{c||}{Outer wheel} & Inner wheel \\
\hline 
 Module & ECC0 & ECC1 & ECC5 & ECC5 \\
\hline 
$<E>$ & 118.8 GeV & 119.1 GeV & 119.3 GeV & 119.1 GeV \\
\hline 
{\bf RMS/}$\mathbf{<E>}$ & {\bf 0.59~\%} &  {\bf 0.52~\%} & {\bf 0.57~\%} & {\bf 0.60~\%}  \\
\hline 
$\sigma/\mu$ & 1.27~\%\ & 1.28~\%\ & 1.22~\%\ & 1.26~\% \\
\hline
{\bf Constant Term} & {\bf 0.70~\%} &  {\bf 0.72~\%} & {\bf 0.61~\%} & {\bf 0.78~\%}  \\
\hline 
\end{tabular}

\vspace*{0.5cm}
\caption{\it Mean energy and non-uniformity of the three tested
modules over the whole analysis region.  For the outer (inner) wheel,
statistical errors on the mean energy are $\sim$0.02~GeV (0.1~GeV) and
statistical errors on non-uniformity are $\sim$0.02~\% (0.1~\%). \label{tab:unifEC}}
\end{center}
\end{table}
\vspace*{0.5cm}


\begin{figure}[htbp]
\begin{center}
\hspace*{-.6cm}
\includegraphics[width=16cm]{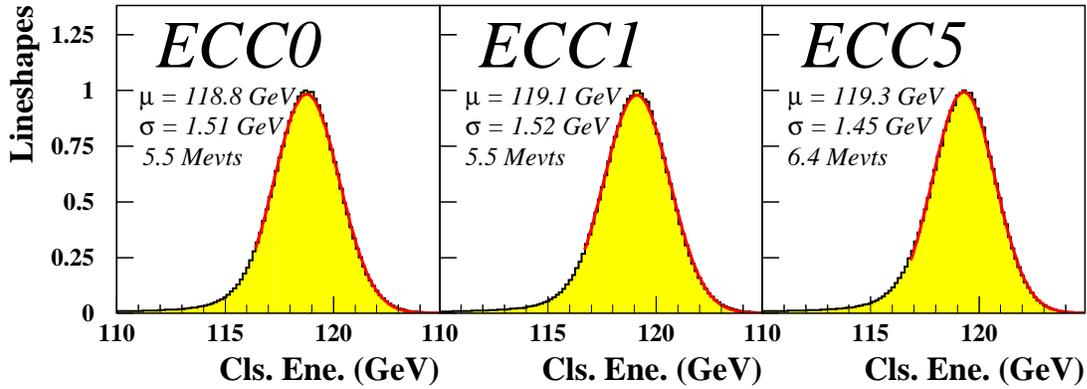}
\caption{\it The energy lineshapes for the endcap modules ECC0, ECC1 and
ECC5 containing respectively 5.5, 5.5 and 6.4 million events . The
simple gaussian fits to energy peak are displayed and the fit
parameters are indicated.}
\label{fig:endcap_lineshapes}
\end{center}
\end{figure}

The overall constant term of the energy resolution is estimated from
the cumulative energy distribution of all cells of the analysis
domain. These energy lineshapes are illustrated in
Fig.~\ref{fig:endcap_lineshapes} for all tested modules. Each of these
energy spectra is fitted with a simple Gaussian form starting from
-1.5$\sigma$ off the mean value. When unfolding a sampling term of
$\sim$11.4$\pm$0.3~GeV$^{1/2}$ an electronic noise term of
$\sim$200~MeV and the beam spread contribution amounting to
$\sim$0.07~\% an overal constant term of $\sim$0.7~\%\ is found. The
global constant terms for all endcap modules are reported in
Table~\ref{tab:unifEC}. 

\section*{CONCLUSION}
\label{Sec:conclu}

\begin{figure}[htbp]
\begin{center}
\includegraphics[width=14cm]{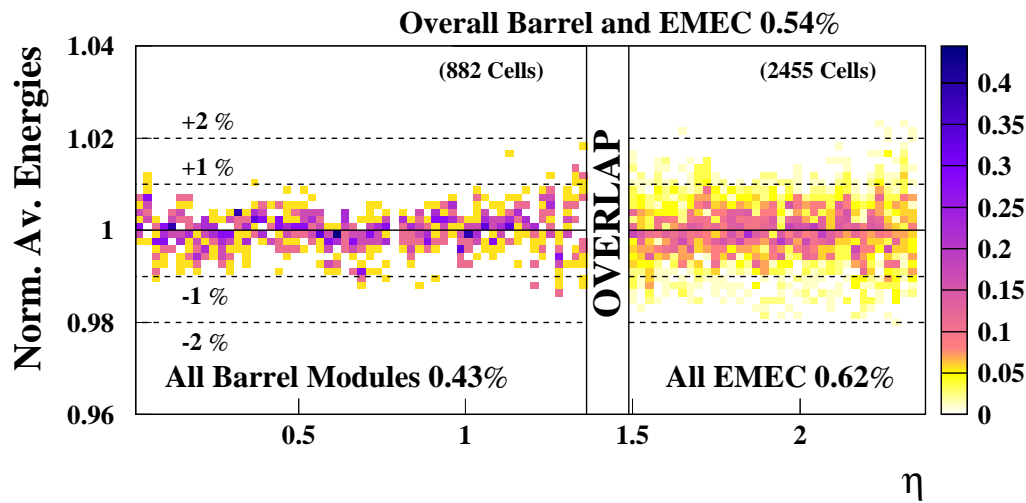}
\caption{\it Two dimensional histogram of the average energies
measured in all cells of all tested modules normalized to the mean
energy of the modules. In the barrel the energies were $\sim$245~GeV
and $\sim$120~GeV in the EMEC. The distributions are normalized to the
number of middle cells scanned in $\phi$ for each value of
$\eta$. \label{fig:summary}}
\end{center}
\end{figure}

The response uniformity of the ATLAS liquid argon electromagnetic
calorimeter to high energy electrons has been studied in the pseudo
rapidity range from 0 to 3.2. These results encompass both the barrel
and endcap calorimeters which were independently tested in different
beam lines using 245 and 119~GeV electrons respectively. The
uniformity in the pseudo rapidity range from 0 to 2.4 is illustrated
in Fig.~\ref{fig:summary}.

For the barrel modules a modified version of the energy reconstruction
scheme developped in~\cite{linearity} mostly based on a full Monte
Carlo simulation is used. Most potential sources of non-uniformity
were reviewed and their impact was independently estimated. When
comparing the estimated non-uniformity to the measured one a very good
agreement is observed thus indicating that the sources of
non-uniformity are well understood. For the endcap modules the
material upstream of the calorimeter is approximately constant, a
simpler energy reconstruction scheme is thus applied. A full Monte
Carlo simulation accounting for the complex geometry of EMEC modules
is performed and is in good agreement with the data.

Non uniformities of the response do not exceed 7~\promille. Overall
constant terms in the energy resolution are derived and range between
5~\promille\ and 7~\promille. Such performance is within the
calorimeter design expectations.

\section*{ACKNOWLEDGMENTS}

We would like to thank the accelerator division for the good working
conditions in the H8 and H6 beam lines. We are indebted to our
technicians for their contribution to the construction and running of
all modules. Those of us from non-member states wish to thank CERN for
its hospitality. 


\end{document}